# An efficient Moving Morphable Component (MMC)-based approach for multi-resolution topology optimization

Chang Liu[1], Yichao Zhu[1], Zhi Sun[1], Dingding Li[1], Zongliang Du[1,2]*, Weisheng Zhang[1], Xu Guo[1]*

[1]*State Key Laboratory of Structural Analysis for Industrial Equipment,*
*Department of Engineering Mechanics,*
*International Research Center for Computational Mechanics,*
*Dalian University of Technology, Dalian, 116023, P.R. China*

[2]*Structural Engineering Department,*
*University of California, San Diego, San Diego, CA 92093, United States of America*

**Abstract**

In the present work, a highly efficient Moving Morphable Component (MMC) based approach for multi-resolution topology optimization is proposed. In this approach, high-resolution optimization results can be obtained with much less numbers of degrees of freedoms (DOFs) and design variables since the topology optimization model and the finite element analysis model are totally decoupled in the MMC-based problem formulation. This is achieved by introducing hyper-elements for structural response analysis and adopting a design domain partitioning strategy to preserve the topological complexity of optimized structures. Both two-and three-dimensional numerical results demonstrate that substantial computational efforts can be saved for large-scale topology optimization problems with the use of the proposed approach.

*Corresponding authors. E-mail: zldu@mail.dlut.edu.cn (Zongliang Du), guoxu@dlut.edu.cn (Xu Guo)

# 1. Introduction

Structural topology optimization, which aims at distributing a certain amount of available materials within a prescribed design domain appropriately in order to achieve optimized structural performances, has been extended to a wide range of physical disciplines such as acoustics, electromagnetics, and optics since the pioneering work of Bendsøe and Kikuchi (1988). So far, classical topology optimization methods have already been implemented in commercial softwares (e.g., Altair-OptiStruct (HyperWorks, 2013) and Abaqus (Simulia, 2011)) to solve practical problems. However, due to the large computational efforts associated with the solution of topology optimization problems, where systems of (sometimes nonlinear) partial differential equations must be solved iteratively to find the structural responses and sensitivity information, topology optimization methods are not easy to be applied to large-scale problems especially when high-resolution designs containing structural features with small length scales are sought for.

In traditional implicit topology optimization methods (e.g., the Solid Isotropic Material with Penalization (SIMP) method, the level set method (LSM)), the finite element analysis (FEA) model and the topology description model are strongly coupled. This means that the density of the FE mesh determines not only the accuracy of FEA, but also the resolution of the obtained optimized solutions. Under this circumstance, very fine FE meshes must be employed if high-resolution designs containing structural features with very small length scales are sought for. This will inevitably lead to large-scale and time consuming computational tasks especially for three-dimensional (3D) topology optimization problems. For example, if a cubic design domain is discretized into $100\times 100\times 100$ elements along three coordinate directions, a FE model with 3 million degrees of freedoms (DOFs) as well as a nonlinear optimization problem with 1 million design variables must be dealt with at every step of the iterative solution process. Furthermore, if we intend to double the resolution to retain more tiny structural features in the optimized design, the corresponding numbers of the DOFs and design variables would increase to 24 million and 8 million, respectively! Recently, with the use of a supercomputer with 8000 processors, Aage et al. (2017) found the optimal reinforcement of a full aircraft wing with 1.1 billion voxels for FE discretization via SIMP method in several days. This, however, is almost an impossible task for ordinary computers.

To promote the practical application of topology optimization, many attempts have been made to enhance the solution efficiency of large-scale problems. One direct approach is to use high performance computers and parallelize the solution process. To be specific, early research works mainly focused on how to obtain structural responses rapidly with use of parallelization techniques (Borrvall and Petersson, 2011; Kim et al., 2004; Vemaganti and Lawrence, 2005; Evgrafov et al., 2007; Mahdavi et al., 2006; Aage et al., 2007). Moreover, in order to reduce the computational time associated with the solution of large-scale nonlinear optimization problems with a huge number of design variables, Aage and Lazarov (2013) also parallelized the well-known MMA optimizer successfully. Although these achievements greatly enhanced the capability of solving large-scale topology optimization problems, the corresponding computational complexity is not reduced essentially. Besides resorting to high performance computing (HPC) techniques, some researchers have also made attempts to enhance the efficiency of FEA by employing some special solution schemes or reducing the total number of DOFs in FEA models directly. For example, Wang et al. (2007) proposed to recycle parts of the search space in a Krylov subspace solver to reduce the number of iterations for solving the equilibrium equations, and significant saving of computational effort is observed especially when the changes of design variables between two consecutive optimization steps are small enough. Amir et al. (2009a) proposed a solution procedure in which exact FEA is performed only at certain stages of iterations while approximate reanalysis is used elsewhere; in Amir et al. (2009b), an alternative stopping criterion for a Preconditioned Conjugate Gradient (PCG) iterative solver was adopted so that fewer iterations are required for obtaining a converged solution. Amir and Sigmund (2010) also proposed an approximate approach to solve the nested analysis equations, and it was reported that the computational cost can be reduced by one order of magnitude. It should be pointed out, however, that the above techniques are generally suitable for dealing with some specific classes of problems and need careful elaborations for more general applications.

Form the aspect of simplifying the FEA model, adaptive mesh refinement techniques (Kim et al., 2003; Stainko, 2005; Guest and Smith Genut, 2010) and model reduction method (Yoon, 2010) have also been introduced in the implicit SIMP-based solution framework. More recently, Nguyen et al. (2009) proposed a multi-resolution formulation for minimum compliance designs based on a coarse FE mesh for structural response analysis and a finer mesh for density field discretization. This treatment can greatly improve the computational efficiency of SIMP-based topology optimization method by reducing the FEA cost. Later on, this approach had been further extended to involve an adaptive mesh refinement scheme (Nguyen et al., 2012). It should be noted that, in the original SIMP-based multi-resolution topology optimization approach (Nguyen et al., 2009), quadrilateral finite elements are adopted for structural analysis. Under this circumstance, in order to obtain meaningful designs with well-connected material distribution, the filter radius has to be comparable to the characteristic size of the adopted FE mesh (not the size of density elements!). As a result, the optimized designs are often suffered from blurred boundaries and may not contain structural details with small feature sizes, although high-resolution density meshes are employed. Nevertheless, recent works (Nguyen et al., 2017; Groen et al., 2017) have shown that, such drawback can be overcome by introducing higher-order finite elements for structural analysis and advanced filter techniques (Guest et al., 2004; Sigmund, 2007; Xu et al., 2010 and Wang et al., 2011). In those contributions, optimized designs with fine structural features and distinct boundaries are obtained successfully in SIMP-based multi-resolution topology optimization framework. However, besides the higher computational effort associated with FEA using higher-order elements, as disclosed by Groen et al. (2017), the order of finite element interpolation needs to be compatible with the resolution ratio between the mesh for density interpolation and the mesh for displacement interpolation, in order to circumvent the issue of artificially stiff patterns. Besides, although the number of DOFs in FEA models can be greatly reduced in SIMP-based multi-resolution topology optimization framework, the number of design variables is still very large in the aforementioned approaches. This, as will be shown later, also leads to a large amount of computational time (corresponding to the solution of large-scale nonlinear/non-convex optimization problems) when large-scale multi-resolution topology optimization problems are considered. In addition, due to the implicit nature of geometry description, post-processing is always required to transfer the optimized designs obtained by implicit topology optimization approaches to computer aided design/engineering (CAD/CAE) systems. This issue to some extent restricts the application of the aforementioned multi-resolution topology optimization approach to large-scale problems which often lead to very complex post-processing works.

In order to overcome the aforementioned challenging issues for solving large-scale multi-resolution optimization problems, in the present paper, the Moving Morphable Components (MMC) based topology optimization approach is extended to the multi-resolution framework. The MMC-based topology optimization method was first initialized in Guo et al. (2014), where a number of structural components with explicit geometry descriptions are adopted as basic building blocks of optimization (see in Fig. 1 for reference). Therefore, optimized designs can be determined by optimizing the explicit geometry parameters characterizing the sizes, shapes and layouts of the introduced components. Compared with traditional topology optimization approaches, in the MMC method, topology optimization actually can be achieved in an explicit and geometrical way. It has been shown that this new solution framework not only can reduce the number of design variables substantially but also has the merit of easily controlling the structural geometry features such as minimum length scale (Zhang et al., 2016a), overhang angle (Guo et al., 2017) and the connectivity of a structure (Deng and Chen, 2016) in an explicit and flexible way. Actually, recent years witnessed a growing interest on developing topology optimization methods based on explicit geometry/topology descriptions (Guo et al., 2016; Zhang et al., 2016b, 2016c, 2017a, 2017b, 2017c, 2018a, 2018b; Liu et al., 2017; Xue et al., 2017; Norato et al., 2015; Zhang et al., 2016, 2017d, 2018; Zhang and Norato, 2017; Zhang et al., 2017e, 2017f; Hoang and Jang, 2017; Hou et al., 2017; Takalloozadeh and Yoon, 2017; Sun et al., 2018 and Xie et al., 2018).

As pointed in Guo et al. (2014) and Zhang et al. (2016c), one of the distinctive features of the MMC-based topology optimization framework is that the corresponding FEA model and the topology description model are *totally decoupled*. In previous implementation of the MMC-based approaches (Guo et al., 2014, 2016, 2017; Zhang et al., 2016a, 2016b, 2016c, 2017b, 2017c, 2018a; Liu et al., 2017; Xue et al., 2017), since the same mesh is used for both the interpolation of displacement field and the projection of explicit structural geometry, the unique decoupling advantage pertaining to the MMC method has not been fully utilized. In the present work, we propose to adopt two sets of meshes with different resolutions for FEA and topology description, respectively, to establish a highly efficient multi-resolution MMC-based solution framework for structural topology optimization. Actually, as will be shown in the forthcoming sections, compared with traditional methods, under the proposed MMC-based multi-resolution framework, with the use of the same linear finite element, the computation time for FEA can be reduced by one order of magnitude, and more importantly, high-resolution designs can be obtained with quite small number of design variables.

The rest of the paper is organized as follows. In Section 2, the problem formulation under the MMC-based solution framework is presented. Then the strategy for obtaining high-resolution designs efficiently using MMCs as basic building blocks of optimization is described in Section 3. Afterwards, some techniques, that are capable of improving the efficiency of numerical implementation of the proposed MMC-based approach and preserving the complexity of structural topology, are introduced in Section 4. In Section 5, several representative examples are presented to illustrate the effectiveness of the proposed approach. Finally, some concluding remarks are provided in Section 6.

## 2. Problem formulation

In the MMC-based topology optimization approach, the material distribution of a structure can be described by a so-called topology description functions (TDF) in the following form:

$$\begin{cases} \phi^s(x) > 0, \text{if } x \in \Omega^s, \\ \phi^s(x) = 0, \text{if } x \in \partial\Omega^s, \\ \phi^s(x) < 0, \text{if } x \in D\backslash(\Omega^s \cup \partial\Omega^s), \end{cases} \tag{2.1}$$

where D represents a prescribed design domain and $\Omega^s \subset D$ denotes the region constituted by $n$ components made of the solid material. As shown in (Guo et al., 2014), the TDF of the whole structure can be constructed as $\phi^s(x) = \max(\phi_1(x), \cdots, \phi_n(x))$ with $\phi_i(x)$ denoting the TDF of the $i$-th component (see Fig. 1 for a schematic illustration). In the present work, for two-dimensional (2D) case, as shown in Fig. 2, $\phi_i(x)$ is constructed as:

$$\phi_i(x,y) = 1 - \left(\frac{x'}{a_i}\right)^p - \left(\frac{y'}{b_i(x')}\right)^p, \tag{2.2}$$

with

$$\begin{Bmatrix} x' \\ y' \end{Bmatrix} = \begin{bmatrix} \cos\theta_i & \sin\theta_i \\ -\sin\theta_i & \cos\theta_i \end{bmatrix} \begin{Bmatrix} x - x_{0i} \\ y - y_{0i} \end{Bmatrix}, \tag{2.3}$$

and $p$ is a relatively large even integer ($p = 6$ in this work). In Eq. (2.2) and Eq. (2.3), the symbols $a_i$, $b_i(x')$, $(x_{0i}, y_{0i})^\top$ and $\theta_i$ denote the half-length, the variable half width, the vector of coordinates of the center and the inclined angle (measured from the horizontal axis anti-clockwisely) of the $i$-th component (see in Fig. 2 for reference), respectively. It should be noted that the variation of the width of the component $b_i(x')$ can take different forms (Zhang et al., 2016c), and in this work it is chosen as

$$b_i(x') = \frac{t_i^1 + t_i^2}{2} + \frac{t_i^2 - t_i^1}{2a_i} x', \tag{2.4}$$

where $t_i^1$ and $t_i^2$ are parameters used to describe the thicknesses of the component.

For 3D case, we use the following TDF to characterize the region occupied by the $i$-th component:

$$\phi_i(x,y,z) = 1 - \left(\frac{x'}{L_i^1}\right)^p - \left(\frac{y'}{h_i(x')}\right)^p - \left(\frac{z'}{f_i(x',y')}\right)^p, \tag{2.5}$$

with

$$\begin{Bmatrix} x' \\ y' \\ z' \end{Bmatrix} = \begin{bmatrix} R_{11} & R_{12} & R_{13} \\ R_{21} & R_{22} & R_{23} \\ R_{31} & R_{32} & R_{33} \end{bmatrix} \begin{Bmatrix} x - x_{0i} \\ y - y_{0i} \\ z - z_{0i} \end{Bmatrix}, \tag{2.6}$$

and

$$\begin{bmatrix} R_{11} & R_{12} & R_{13} \\ R_{21} & R_{22} & R_{23} \\ R_{31} & R_{32} & R_{33} \end{bmatrix} = \begin{bmatrix} c_b \cdot c_t & -c_b \cdot s_t & s_b \\ s_a \cdot s_b \cdot c_t + c_a \cdot s_t & -s_a \cdot s_b \cdot s_t + c_a \cdot c_t & -s_a \cdot c_b \\ -c_a \cdot s_b \cdot c_t + s_a \cdot s_t & c_a \cdot s_b \cdot s_t + s_a \cdot c_t & c_a \cdot c_b \end{bmatrix}, \tag{2.7}$$

respectively. In Eq. (2.7), $s_a = \sin\alpha$, $s_b = \sin\beta$, $s_t = \sin\theta$, $c_a = \sqrt{1 - s_a^2}$, $c_b = \sqrt{1 - s_b^2}$ and $c_t = \sqrt{1 - s_t^2}$ with $\alpha, \beta$ and $\theta$ denoting the rotation angles of the component from a global coordinate system $Oxyz$ to the local coordinate system $O'x'y'z'$, respectively (see Fig. 3 for reference). The vector of the coordinates of the central point and the half-length of the component are represented by the coordinate $(x_{0i}, y_{0i}, z_{0i})^\top$ and $L_i^1$, respectively. Furthermore, the functions $h_i(x')$ and $f_i(x', y')$ in Eq. (2.5) are used to describe the thickness profiles of the component in $y$ and $z$ directions, respectively. In this work, $h_i(x')$ and $f_i(x', y')$ are simply chosen as

$$h_i(x') = L_i^2, \quad f_i(x', y') = L_i^3, \tag{2.8}$$

as shown in Fig. 4. Other forms of $h_i(x')$ and $f_i(x', y')$ can be found in (Zhang et al., 2017c).

With use of the above expressions, the region $\Omega_i^s$ occupied by the $i$-th component can be described as:

$$\begin{cases} \phi_i(\mathbf{x}) > 0, \text{if } \mathbf{x} \in \Omega_i^s, \\ \phi_i(\mathbf{x}) = 0, \text{if } \mathbf{x} \in \partial\Omega_i^s, \\ \phi_i(\mathbf{x}) < 0, \text{if } \mathbf{x} \in D\setminus(\Omega_i^s \cup \partial\Omega_i^s). \end{cases} \tag{2.9}$$

It is also obvious that $\Omega^s = \cup_{i=1}^n \Omega_i^s$. At this position, it is worth noting that topology optimization can also be carried out in the MMC-based solution framework without introducing TDF. Actually, the TDF is only employed for the convenience of performing

FEA under fixed mesh. We refer the readers to (Zhang et al., 2017a, 2018b) for the implementation of the MMC-based topology optimization approach without using TDFs.

Based on the above description, it is obvious that the layout of a structure can be solely determined by $\boldsymbol{D} = \left((\boldsymbol{D}^1)^\top, \ldots, (\boldsymbol{D}^i)^\top, \ldots (\boldsymbol{D}^n)^\top\right)^\top$, a vector of design variables. To be specific, for 2D case, we have $\boldsymbol{D}^i = (x_{0i}, y_{0i}, a_i, \boldsymbol{d}_i^\top, \theta_i)^\top$, which contains the design variables associated with the $i$-th component with $\boldsymbol{d}_i$ denoting the vector of geometry parameters related to $b_i(x')$. In 3D case, $\boldsymbol{D}$ can also be constructed in a similar way.

Based on the above descriptions, a typical topology optimization problem under the MMC-based solution framework can be formulated as follows:

$$\text{Find } \boldsymbol{D} = \left((\boldsymbol{D}^1)^\top, \ldots, (\boldsymbol{D}^i)^\top, \ldots, (\boldsymbol{D}^n)^\top\right)^\top$$

$$\text{Minimize } I = I(\boldsymbol{D})$$

$$\text{S.t.} \tag{2.10}$$

$$g_k(\boldsymbol{D}) \leq 0, k = 1, \ldots, m,$$

$$\boldsymbol{D} \subset \mathcal{U}_{\boldsymbol{D}},$$

where $I(\boldsymbol{D})$, $g_k$, $k = 1, \ldots, m$ are the objective function/functional and constraint functions/functionals. In Eq. (2.10), $\mathcal{U}_{\boldsymbol{D}}$ is the admissible set that design variable vector $\boldsymbol{D}$ belongs to.

In the present study, structures are designed to minimize the structural compliance under the volume constraint of available solid material. Under this circumstance, the corresponding problem formulation can be specified as:

$$\text{Find } \boldsymbol{D} = \left((\boldsymbol{D}^1)^\top, \ldots, (\boldsymbol{D}^i)^\top, \ldots, (\boldsymbol{D}^n)^\top\right)^\top, \boldsymbol{u}(\boldsymbol{x}) \in H^1(\Omega^s)$$

$$\text{Minimize } C = \int_D H(\phi^s(\boldsymbol{x};\boldsymbol{D}))\boldsymbol{f} \cdot \boldsymbol{u}\,\mathrm{d}V + \int_{\Gamma_t} \boldsymbol{t} \cdot \boldsymbol{u}\,\mathrm{d}S$$

$$\text{S.t.} \tag{2.11}$$

$$\int_D H^q(\phi^s(\boldsymbol{x};\boldsymbol{D}))\mathbb{E}\colon \boldsymbol{\varepsilon}(\boldsymbol{u})\colon \boldsymbol{\varepsilon}(\boldsymbol{v})\,\mathrm{d}V = \int_D H(\phi^s(\boldsymbol{x};\boldsymbol{D}))\boldsymbol{f} \cdot \boldsymbol{v}\,\mathrm{d}V$$

$$+ \int_{\Gamma_t} \boldsymbol{t} \cdot \boldsymbol{v}\,\mathrm{d}S, \quad \forall \boldsymbol{v} \in \mathcal{U}_{\text{ad}},$$

$$\int_D H(\phi^s(\boldsymbol{x};\boldsymbol{D}))\,\mathrm{d}V \leq \bar{V},$$

$$\boldsymbol{D} \subset \mathcal{U}_{\boldsymbol{D}},$$

$$\boldsymbol{u} = \bar{\boldsymbol{u}}, \text{on } \Gamma_u,$$

where D, $\boldsymbol{f}$, $\boldsymbol{t}$, $\boldsymbol{u}$, $\boldsymbol{\varepsilon} = \text{sym}(\nabla \boldsymbol{u})$ and $\bar{\boldsymbol{u}}$ are the design domain, the body force density, the prescribed surface traction on Neumann boundary $\Gamma_t$, the displacement field, the linear strain tensor and the prescribed displacement on Dirichlet boundary $\Gamma_u$, respectively. The symbol $H = H(x)$ denotes the Heaviside function with $H = 1$ if $x > 0$ and $H = 0$ otherwise. For numerical implementation purpose, $H(x)$ is often replaced by its regularized version $H_\epsilon(x)$. In the present work, $H_\epsilon(x)$ is taken as

$$H_\epsilon(x) = \begin{cases} 1, & \text{if } x > \epsilon, \\ \frac{3(1-\alpha)}{4}\left(\frac{x}{\epsilon} - \frac{x^3}{3\epsilon^3}\right) + \frac{1+\alpha}{2}, & \text{if } -\epsilon \leq x \leq \epsilon, \\ \alpha, & \text{otherwise,} \end{cases}$$

(2.12)

where $\epsilon$ and $\alpha$ are two small positive numbers used for controlling the length of the transition zone for the regularization of $H(x)$ and avoiding the singularity of the global stiffness matrix, respectively. In Eq. (2.11), $\phi^s(\boldsymbol{x}; \boldsymbol{D})$ is the TDF of the whole structure while $q > 1$ is a penalization factor (in the present work, $q = 2$ is used). In Eq. (2.11), $\mathbb{E} = E^s/(1+\nu^s)[\mathbb{I} + \nu^s/(1-2\nu^s)\boldsymbol{\delta} \otimes \boldsymbol{\delta}]$ is the fourth order elasticity tensor of the isotropic solid material with $E^s$, $\nu^s$, $\mathbb{I}$ and $\boldsymbol{\delta}$ denoting the Young's modulus as well as the Poisson's ratio of the solid material, the symmetric part of the fourth order identity tensor and the second order identity tensor, respectively. The symbol $\mathcal{U}_{ad} = \{\boldsymbol{v} | \boldsymbol{v} \in \boldsymbol{H}^1(\Omega^s), \boldsymbol{v} = \boldsymbol{0} \text{ on } S_u\}$ represents the admissible set of virtual displacement vector $\boldsymbol{v}$ and $\bar{V}$ is the upper limit of the volume of the available solid material.

## 3. Solution strategies for multi-resolution topology optimization under the MMC-based framework

In this work, the hyper-element technique proposed by Nguyen et al. (2009) is adopted in the MMC-based solution framework to construct a highly efficient multi-resolution topology optimization approach. As described in (Nguyen et al., 2009), the basic idea of the hyper-element-based approach is that two sets of meshes with different resolutions are used for solving a topology optimization problem (see Fig. 5 for reference). The coarse meshes are used for interpolating the displacement field while the refined background elements are used for describing the structural geometry with high resolution. This method has been proven to be very effective to reduce the computational cost associated with FEA in SIMP method. It should be noted that, however, that when quadrilateral finite elements are adopted for structural analysis (Nguyen et al., 2009), the filter radius has to be comparable with the size of FE mesh to avoid the checkboard pattern in optimization results. As a result, the optimized structures always do not contain structural details with small feature sizes although high-resolution density meshes are employed.

However, if such hyper-element technique is applied under the MMC-based solution framework, the situation is totally different. This is because in the MMC approach, the structural geometry is described by a set of explicit geometrical parameters. This means that, theoretically speaking, the structural topology has an *infinitely high resolution* in the MMC approach. Based on this consideration, in the present work, we propose to combine both the advantages of the hyper-element approach and the MMC-based solution framework to tackle the multi-resolution topology optimization problems in a computationally efficient way.

In the present work, as the same in traditional treatments, we also intend to use a fixed FE mesh and an ersatz material model for FEA, although adaptive FE mesh can also be applied to calculate structural responses since we have the explicit boundary representation in the MMC approach. Under this circumstance, refined background elements are also needed to identify the small structural features. It is, however, worth noting that, as can be seen clearly from the following discussions, unlike the traditional implicit topology optimization method, the refinement of the background elements does not increase the number of design variables, and it only increases the computational effort associated with numerical integrations when the element stiffness matrix is calculated.

With the use of the hyper-element technique, the stiffness matrix of the $i$-th hyper-element can be calculated as (see Fig. 6 for a schematic illustration):

$$\boldsymbol{K}_i = \int_{\Omega_i} \boldsymbol{B}^\top \boldsymbol{D}_i(\boldsymbol{x}) \boldsymbol{B} \mathrm{d}\Omega \approx \sum_{j=1}^{ng} E_{i,j} \boldsymbol{B}(\boldsymbol{x}_{i,j}^0)^\top \boldsymbol{D}_0 \boldsymbol{B}(\boldsymbol{x}_{i,j}^0) A_g, \quad (3.1)$$

where $\Omega_i$ represents the region occupied by the $i$-th hyper-element, $\boldsymbol{x} = (x, y)$ is the vector of spatial coordinates, $\boldsymbol{B}$ and $\boldsymbol{D}_i$ are the strain-displacement matrix and the constitutive matrix, respectively. In Eq. (3.1), $ng$ represents the number of background elements in the considered hyper-element, $\boldsymbol{D}_0$ corresponds to the constitutive matrix of the solid material with unit Young's modulus and $A_g$ is the area of a background element, respectively. $\boldsymbol{x}_{i,j}^0$ is the coordinate vector of the integration point (simply chosen as the central point of the corresponding background element in the present work) associated with the $j$-th background element in the $i$-th hyper-element. In addition, $E_{i,j}$ is the smeared Young's modulus of the $j$-th background element in the $i$-th hyper-element. Under the spirit of the ersatz material model, $E_{i,j}$ can be calculated through the corresponding nodal values of the TDF as

$$E_{i,j}(\phi^s) = \frac{E^s \left(\sum_{e=1}^{4} \left(H(\phi_{i,j}^{se})\right)^q\right)}{4}, \tag{3.2}$$

where $\phi_{i,j}^{se}$ is the value of TDF of the whole structure at the $e$-th node of element $(i,j)$, $E^s$ is the Young's modulus of the solid material.

Once the element stiffness matrix of each *hyper-element* is obtained, we can then assemble the global stiffness matrix $\boldsymbol{K}$, solve the displacement vector $\boldsymbol{U}$ and obtain the structural compliance as $C = \boldsymbol{U}^\top \boldsymbol{K} \boldsymbol{U} = \sum_{i=1}^{NS} \boldsymbol{U}_i^\top \boldsymbol{K}_i \boldsymbol{U}_i$ with $\boldsymbol{U}_i$ denoting the nodal displacement vector of the $i$-th hyper-element and $NS$ representing the total number of hyper-elements. Then the sensitivity of the structural mean compliance with respect to a design variable $d$ (in the context of FEA) can be expressed as:

$$\frac{\partial C}{\partial d} = -\sum_{i=1}^{NS} \boldsymbol{U}_i^\top \frac{\partial \boldsymbol{K}_i}{\partial d} \boldsymbol{U}_i$$

$$= -\sum_{i=1}^{NS} \boldsymbol{U}_i^\top \left( \frac{E_0}{4} \sum_{j=1}^{ng} \left( \sum_{e=1}^{4} q \left(H(\phi_{i,j}^{se})\right)^{q-1} \frac{\partial H(\phi_{i,j}^{se})}{\partial d} \right) \boldsymbol{B}(\boldsymbol{x}_{i,j}^0)^\top \boldsymbol{D}_0 \boldsymbol{B}(\boldsymbol{x}_{i,j}^0) A_g \right) \boldsymbol{U}_i. \tag{3.3}$$

For the volume constraint, we also have

$$\frac{\partial V}{\partial d} = \frac{1}{4} \sum_{j=1}^{NG} \sum_{e=1}^{4} \frac{\partial H(\phi_j^{se})}{\partial d}. \tag{3.4}$$

The derivation of $\partial H(\phi_j^{se})/\partial d$ in Eq. (3.3) and Eq. (3.4) is trivial and will not be repeated here.

## 4. Numerical implementation aspects

In this section, we will discuss some numerical techniques that will be used to implement the proposed MMC-based multi-resolution topology optimization approach in a computationally efficient way. Actually, these techniques are not only applicable to the multi-resolution design case, but also capable of enhancing the computational efficiency of the original single-resolution oriented MMC approach (Guo et al., 2014; Zhang et al., 2016c). Moreover, a so-called design domain partitioning strategy is developed to preserve the topological complexity of the optimized designs obtained by the proposed multi-resolution topology optimization approach.

*4.1 Generating the TDF of the structure and calculating sensitivities locally*

As shown in Section 2, the geometry of a component is described by a $p$-th order hyperelliptic function. In our previous numerical implementations (e.g., Zhang et al., 2016c), the TDF values associated with each component are calculated at every node of background FE mesh with use of Eq. (2.2)-Eq. (2.4) (for 2D case) or Eq. (2.5)-Eq. (2.8) (for 3D case). If, for example, a problem with 500 components and $1000 \times 500$ background elements is considered, the TDF nodal values must be calculated $(1000+1) \times (500+1) \times 500$ times to generate the TDF of the whole structure. This treatment will definitely consume a large amount of computational time and computer memory and therefore is not suitable for solving large scale problems.

Actually, the nodal TDF values of the background FE mesh are used for the following three purposes: 1) describing the geometry of the components through Eq. (2.9); 2) calculating the Heaviside function used in the ersatz material model using Eq. (2.12) and 3) carrying out the sensitivity analysis as shown in Eq. (3.3)-Eq. (3.4). Actually, a component only occupies a small portion of the design domain, so it is not necessary to calculate the nodal values of TDF on the whole region. Furthermore, from Eq. (2.12), it can also be observed that both the regularized Heaviside function and its derivative with respect to the TDF only vary in a narrow band $\Omega^{\mathrm{BDY}} = \{x | x \in \mathrm{D}, -\epsilon \leq \phi^{\mathrm{s}}(x) \leq \epsilon\}$ around the structural boundary and keep constant in rest of the design domain. These observations inspire us that we can only generate and store the nodal values of the TDF of each component around its boundary *locally* (see Appendix for more details). Since the size of an individual component is usually relative small compared with that of the whole design domain, this strategy can save the computational effort and computer memory used to generate the corresponding TDF significantly.

In previous numerical implementation (e.g., Zhang et al., 2016c), the formula $\phi^{\mathrm{s}} = \max(\phi_1, \phi_2, \cdots, \phi_n)$ is used to generate the TDF of whole structure. In the present work, the following well-known K-S function is used to approximate the max operation (Kreisselmeier and Steinhauser, 1979):

$$\phi^s \approx \ln\left(\sum_{i=1}^{n} \exp(l\phi_i)\right)/l, \tag{4.1}$$

where $l$ is a large positive number (e.g., $l = 100$). Using the same method mentioned before, the exponent arithmetic in Eq. (4.1) can be carried out only around the boundary region of each component (see more details in Appendix). Numerical experiments indicate that this treatment can also enhance the computational efficiency of generating the TDF of whole structure significantly.

In addition, since the regularized Heaviside function's derivative with respect to the TDF only varies near the structural boundary, the sensitivities of the objective and constraint functions also can be calculated locally. It is worth noting that, although the sensitivity analysis in the MMC approach is not as straightforward as that in SIMP approach, the time cost for sensitivity analysis associated with the proposed new implementation of the MMC method with local evaluation is, however, still much less than (or at least comparable to) that in the SIMP approach. This is due to the fact that the number of design variables are significantly reduced and there is no chain rule operation resulting from the non-local filter operator is involved in the present MMC-based approach. This point will be verified by the numerical examples provided in Section 5.

*4.2 Design domain partitioning strategy for preserving structural complexity*

In this subsection, we shall discuss how to control the topological complexity in optimized designs. In the MMC approach, as shown in Fig. 7a, the components can move, morph, disappear, overlap and intersect with each other to generate an optimized structure. Since the sensitivities are nonzero only in a narrow band near the structural boundary, it is not difficult to observe that the sensitivities of the objective/constraint functions with respect to the design variables associated with a hidden component are zero. In other words, once a component is fully covered by other components, the design variables associated with this component will remain unchanged in the following optimization process unless the components cover it move away. Actually, this mechanism is responsible for the relatively simple topology of the optimized designs obtained by MMC approach, since many components may be covered by other components in the final optimized results (see Fig. 7a for reference).

Although a design with simple structural topology may be more favorable from manufacturing point of view, however, theoretical analysis indicated that optimal solutions of topology optimization problems may possess very complex structural topologies (e.g., the Michell truss (Sigmund et al., 2016; Dewhurst, 2001)). As a result, it is very necessary to equip the MMC approach with the capability of producing optimized designs with complex structural topologies.

Actually, the aforementioned goal can be achieved by resorting to the so-called design domain partitioning strategy. The key point is to restrain the range of the motions of the components. As shown in Fig. 7b, in the proposed design domain partitioning strategy, the design domain D is divided into several non-overlapped sub-regions $\Omega_i^{\text{sub}}$, $i = 1, \ldots, ns$, where a specific number of components are distributed in these sub-regions initially. During the entire process of optimization, it is required that the central point of every component initially located in a specific sub-region is always confined in that sub-region. This can be achieved easily by imposing some upper/lower bounds on the coordinates of central points of involved components in the MMC-based problem formulation. In addition, this strategy actually can provide a flexible way to control the structural complexity locally and adaptively. For example, if it is intended to produce an optimized structure with high structural complexity in a specific region $D_\alpha \subset D$, we can divide $D_\alpha$ into a relatively large number of sub-regions $\Omega_{j\alpha}^{\text{sub}}$, $j = 1, \ldots, n_\alpha^s$ (i.e., $D_\alpha = \bigcup_{j=1}^{n_\alpha^s} \Omega_{j\alpha}^{\text{sub}}$) and put a relatively large number of components in each $\Omega_{j\alpha}^{\text{sub}}$. With the use of this treatment, it can be expected that the corresponding optimized structure may possess complex structural topology and structural features with length scales comparable with the characteristic sizes of the sub-regions in $D_\alpha$. The effectiveness of this design domain partitioning strategy will be verified numerically in the forthcoming section.

At this position, it is also interesting to note that the proposed solution framework also has some underlying relationship with the classical approaches. This can be explained as follows. Actually, in the proposed method, the sub-regions can be selected as being coincided with the finite elements used for interpolating the displacement field, and only one component is distributed in each sub-region (element) individually in a form as shown in Fig. 8. Furthermore, we can only take the heights of the components as design variables and interpolate the Young's modulus of each element in terms of $h_i$ as $E_i = E^s (h_i/H_i)^p$ with $h_i$ and $H_i$ denoting the heights of the component and the corresponding finite element (sub-region) ($h_i \leq H_i$), respectively. Under the above treatment, it can be observed clearly that the proposed MMC-based multi-resolution topology optimization approach will degenerate to the classical SIMP approach by defining the value of $h_i/H_i$ as the corresponding element density.

## 5. Numerical examples

In this section, three plane stress examples with unit thickness and one 3D example are investigated to illustrate the effectiveness of the proposed MMC-based method for multi-resolution topology optimization. The computational time and the optimized objective function values are compared with their counterparts obtained by efficient implementations of the SIMP method (i.e., 88-lines 2D code in (Andreassen et al., 2010); 169-lines 3D code in (Liu and Tovar, 2014)). In the 2D examples, the MMA algorithm (Svanberg, 1987) is chosen as the optimizer for both the MMC and the SIMP methods. In the 3D example, the Optimality Criteria (Bendsøe, 1995) and MMA algorithms are used in the SIMP and the proposed MMC method, respectively. The termination criteria is satisfied when $\frac{\|c_i - \overline{c^5}_i\|}{\overline{c^5}_i} \leq 5 \times 10^{-4}$, $V_i \leq \bar{V}$ and $\frac{\|V_i - \overline{V^5}_i\|}{\overline{V^5}_i} \leq 5 \times 10^{-4}$, $i = 5, 6, 7, ...$, where $c_i$ and $V_i$ are the objective function value and the volume of solid material in the $i$-th step, $\overline{c^5}_i$ and $\overline{V^5}_i$ are the average value of the objective function and the average volume of solid material in the last five consecutive iteration steps, $\bar{V}$ is the upper bound of the available volume of the solid material. Without loss of generality, all involved quantities are assumed to be dimensionless. The Young's modulus and the Poisson's ratio of the isotropic solid material are chosen as $E^s = 1$ and $v^s = 0.3$, respectively. In addition, all computations are carried out on a Dell-T5810 workstation with an Intel(R) Xeon(R) E5-1630 3.70GHz CPU, 128GB RAM of memory, Windows10 OS, and the computer code is developed in MATLAB 2016b. The values of parameters in Eq. (3.2) and Eq. (2.12) are taken as $q = 2$, $\epsilon = 2 \times \min(\Delta x, \Delta y, \Delta z)$ and $\alpha = 10^{-3}$, respectively, unless otherwise stated. Here $\Delta x, \Delta y$ and $\Delta z$ are the sizes of the *background elements* along three coordinate directions.

*5.1 A cantilever beam example*

In this example the well-known short cantilever beam problem is examined. The design domain, external load, and boundary conditions are all shown in Fig. 9. A 12×6 rectangular design domain is discretized by 1280×640 uniform quadrilateral background elements for geometry representation. A unit vertical load is imposed on the middle point of right boundary of the design domain. The available volume of the solid material is $\bar{V} =$ $0.4V_\mathrm{D}$ with $V_\mathrm{D}$ denoting the volume of the design domain. Fig. 10 shows the initial design composed of 576 components.

Firstly, the effectiveness of the design domain partitioning strategy described in the previous section is examined. To this end, the design domain is divided into $1\times 1$, $6\times 3$, $12\times 6$ sub-regions along the horizontal and vertical directions, respectively. For all cases, $1280 \times 640$ uniform quadrilateral plane stress elements are used for FEA. The corresponding optimized designs are shown in Fig. 11. It is obvious that as the number of sub-regions is increased, the optimized structural topology becomes more complicated, meanwhile the objective function value is slightly decreased. This reflects that the design domain partitioning strategy is very effective to control the topological complexity of the optimized designs.

Next, the number of sub-regions is fixed as $12\times 6$ and the efficiency of the proposed multi-resolution algorithm is further investigated. For different resolutions of hyper-element meshes for FEA while keeping the same number of background elements, the optimized designs, iteration numbers and the average time costs for some key parts of the corresponding optimization process are shown in Table 1. In this table, the parameters $\bar{t}_\mathrm{TDF}$, $\bar{t}_\mathrm{FEA}$, $\bar{t}_\mathrm{sen}$, $\bar{t}_\mathrm{MMA}$ represent the *average* time costs in one optimization step for constructing the TDFs of the components, assembling and solving the FEA equations, sensitivity analysis and MMA optimizer respectively, and $\bar{t}_\mathrm{total}$ represents the *average* time costs of an entire optimization step. The symbol $n_\mathrm{iter}$ represents the final number of iteration when the convergence criterion is satisfied. The quantities $c_\mathrm{obj}$ and $c_\mathrm{post}$ represent the value of the object function obtained with the hyper-element mesh and the background element mesh, respectively. It is found that, as the number of the hyper-elements is gradually reduced, the time cost of FEA decreases rapidly, which shares the same advantage of SIMP-based multi-resolution topology optimization approaches (Nguyen et al., 2009, 2017; Groen et al., 2017). To be specific, for this example, the total number of degree of freedom is 1642242 when the background elements mesh (with a number of $1280\times 640$ elements) are used for structural analysis, while this number decreases to 26082 when $160\times 80$ hyper-elements mesh is used. Accordingly, as shown in Table 1, the average time cost of FEA is decreased sharply from 15.18s to 0.47s per optimization step. It is also found that as the number of the iterations has a slight increase as the number of the hyper-elements is reduced. Some enlarged plots of the optimized structures and the corresponding iteration curves are provided in Fig. 12.

On the other hand, since a coarser FE mesh would overestimate the structural stiffness, the accuracy of the adopted multi-resolution strategy should also be examined seriously. In Table 1, the converged values of the objective function as well as the relative errors of FEA results are provided. When the resolution ratio $n_{\text{be}}$ between the background element mesh and hyper-element mesh is less than or equal to 8, the relative errors are less than 4%. However, as seen in the last two cases, in Table 1, when the resolution ratio is very large (e.g., $n_{\text{be}} \geq 10$), unacceptable FEA error may be introduced. In addition, optimized designs with small voids and even designs with disconnected material distribution may be obtained when low resolution FE meshes are adopted. This is because small voids and disconnected components cannot be identified successfully by very coarse hyper-element mesh by the integration scheme adopted. This will inevitably lead to the overestimation of structural stiffness. This issue can be resolved by using more number of hyper elements or introducing higher order interpolation schemes as in (Nguyen et al., 2017; Groen et al., 2017). Another interesting observation is that, different from the SIMP-based multi-resolution topology optimization results with quadrilateral elements (Nguyen et al., 2009), in the proposed approach, no filter operation is required and structural features with very small characteristic sizes can be well-preserved in the final optimized designs even for coarser hyper-element meshed.

Finally, the optimization results obtained by the proposed method are also compared with those obtained by the 88-line implementation of the SIMP method (Andreassen et al., 2010, with $E_{\text{min}} = 10^{-9}$, penalty factor $p = 3$ and the radius of density filter $r = 1.2$, respectively, see Table 2 for more details) to illustrate the distinctive features of the proposed method. One can be observed that: 1) The proposed MMC method only needs a little time cost for updating the TDFs. 2) Most computational time in the SIMP approach is paid for FEA and updating design variables. For the same FE mesh, the computational time for FEA corresponding to the proposed method and the SIMP method are almost the same. If, however, the hyper-element technique is adopted, structural responses with reasonable accuracy (a relative error less than 5%) can be obtained with much less computational time (about 1/30). Moreover, since the number of design variables in the MMC method is only 3456 (as compared to 819200 in the SIMP approach), the computational efficiency for updating design variables by MMA optimizer in the proposed approach can be improved by more than 200 times compared to that of the SIMP approach (actually 0.05s vs 14.70s!). As a result, when the same FE mesh is used, the average computational time for one optimization step is about 28.91s in the SIMP approach while the value is about 18.62s in the proposed approach, which can be further decreased to 3.13s when the hyper-element technique is employed. This comparison clearly verifies the effectiveness of our method for solving large scale topology optimization problems efficiently. 3) Since no filter operation is applied to eliminate numerical instabilities, the optimized designs obtained by the proposed approach are pure black-and-white and share some features of the classical Michell truss structures. The advantage can be further illustrated by comparing the value of the objective functional. Actually, by adopting the same interpolation strategy for Young's modulus of non-solid elements in the SIMP approach, the value of the objective functional for the optimized design obtained by $1280 \times 640$ FE mesh is 74.72, which is smaller than that (i.e., 80.29) of the design obtained by the SIMP approach.

*5.2 The MBB example*

The setting of this example is described schematically in Fig. 13. A vertical load $f = 2$ is imposed on the middle point of the top side of the beam. For simplicity, only half of the design domain is discretized by a $1280 \times 640$ uniform background element mesh for geometry description. In this example, the upper bound of the volume of available solid material is set to be $\bar{V} = 0.4 V_\mathrm{D}$.

The design domain is divided into $12 \times 6$ equal square sub-regions to preserve the structural complexity. The initial layout of the components is the same as that in the previous cantilever beam example (see Fig. 10 for reference). By interpolating the displacement field with $640 \times 320$, $320 \times 160$, $258 \times 128$ and $160 \times 80$ hyper-elements, respectively, as shown in Table 3, the computational time for FEA can be reduced by almost 25 times as compared with the case where the background element mesh (i.e., $1280 \times 640$) is adopted for FEA. It is found that when $n_\mathrm{be} = 8$, the relative error of the value of objective functional reached to 10.21%, however, the corresponding value of the object functional recalculated by the background element mesh (i.e. 97.86) is still very close to those obtained by smaller resolution ratios (e.g. 96.98 for $n_\mathrm{be} = 5$). From this point of view, the proposed multi-resolution approach is still supposed to be effective for such case. The enlarged figures of the optimized structures and the corresponding iteration histories associated with the case where $640 \times 320$ and $256 \times 128$ hyper-elements are adopted, can be found in Fig. 14. Table 4 provides the optimization result obtained by the SIMP approach under a 1280×640 FE mesh with $E_{\min} = 10^{-9}$, penalty factor $p = 3$ and the radius of density filter $r = 1.2$, respectively. By comparing the corresponding results in Table 3 and Table 4, similar conclusions can be made as in the previous example.

*5.3 A cantilever beam subject to a distributed load*

In this example, a cantilever beam under a uniform distributed load introduced in (Groen et al., 2017) is revisited. The setting of the problem is described schematically in Fig. 15. A vertical distributed load is imposed on the top surface of the design domain uniformly with density $f = 1/l$. During solution process, with the use of 6×3 sub-regions, the design domain is discretized by a 1200×600 uniform background element mesh for geometry representation. The initial design is the same as that in the first cantilever beam example (see Fig. 10 for reference). A detailed discussion about this example can be found in (Groen et al., 2017) by adopting the SIMP-based higher-order multi-resolution topology optimization method.

Firstly, the maximum available solid material volume is set as $\bar{V} = 0.4V_\mathrm{D}$ (the same as that in (Groen et al., 2017)), and 1200×600 uniform quadrilateral plane stress elements are used for FEA. The corresponding iteration histories are plotted in Fig. 16a. It is found that the value of the objective functional oscillates during the optimization prosses. This is because although the structural topology has been already obtained after about 150 iterations, some small voids emerge and disappear alternately in the region around the top surface of the structure (see Fig. 16b-Fig. 16d for reference). To circumvent this unpleasure behavior, the top layer of the background element mesh of the design domain is fixed as solid elements in numerical implementation. The optimized structure obtained under this treatment and corresponding iteration histories can be seen in Fig.17. It is found that more stable convergence history is achieved and the value of the objective functional of the optimized structure is very close to that of the structure shown in Fig. 8a of (Groen et al., 2017).

Next, with the numbers of the background elements (1200×600) and the undesign domain (solid top layer) keep fixed, the effectiveness of the proposed multi-resolution approach is tested by adopting a smaller available volume of solid material $\bar{V} = 0.3V_\mathrm{D}$. The displacement field is discretized by 600×300, 400×200, 300×150, 200×100 and 120×60 hyper-elements, respectively, and the obtained results are summarized in Table

5. It is found that, for this example, when the number of hyper-elements is $300\times 150$, the relative error of the value of the objective functional is 47.34%. As a result, the admissible resolution ratio $n_{\mathrm{be}}$ decreases to 3, which is much less than that in the above examples where concentrated forces are considered. This is because in the FEA model, the distributed load must be translated to nodal forces of the hyper-elements by virtual work principle, and the accuracy of this treatment is directly determined by the resolution of the FE mesh. Nevertheless, it is observed that the maximum admissible resolution ratio can be increased by increasing the number of the undesign layers. For instance, when six top layers of the background element mesh are fixed as solid elements, the maximum resolution ratio $n_{\mathrm{be}}$ can be increased to 6, and the corresponding optimized structures and corresponding iteration curves for the cases where $300\times 150$, $200\times 100$ and $120\times 60$ hyper-elements are used, respectively, can be found in Fig. 18.

*5.4 A 3D box example*

This example is a variation of the one presented in (Sigmund et al., 2016). As illustrated in Fig. 19a, the design domain is a $12\times 10\times 12$ box, which is subjected to a pair of torque. The torque load is simulated by four concentrated point-forces as described in Fig. 19 and the magnitudes of these point forces are chosen as $f=2$. The radii of the two red disks are 1.5 and their thicknesses are 0.15, respectively. Two void parts (the gray cylinder regions in Fig. 19a) are fixed as non-design domains. For simplicity, only 1/8 of the design domain is optimized. The maximum volume fraction of the available solid material is 2%.

This problem is solved with use of the proposed approach for three sets of background element mesh (i.e., $42\times 35\times 42$, $84\times 70\times 84$ and $126\times 105\times 126$, respectively). The same initial design containing 720 components, as shown in Fig. 19b, is adopted for all three tested cases. For comparison, this example is also solved by the SIMP method with use of its efficient numerical implementation described in Liu and Tovar (2014), with $E_{\min}=10^{-9}$, penalty factor $p=3$ and the radius of density filter $r=1.5$, respectively. Optimality criterion (OC) method is used for updating the design variables in the SIMP method. Note that OC method is adopted here for updating the design variables since it is more efficient than the MMA method when large number of design variables are involved. It should be pointed out that, for the current hardware setting, the computer memory (128G) would be run out when $84\times 70\times 84$ traditional 8-node brick elements are used for FEA, which is implemented in MATLAB environment. Therefore, we can only use $42\times35\times42$ FE meshes in the SIMP approach and $42\times35\times42$ *hyper-elements* for all there background element mesh cases in the proposed approach for FEA, respectively.

The entire structure obtained by the SIMP method is shown in Fig. 20. The compliance of the 1/8 optimized structure is 120.49. Since the optimized solution obtained by the SIMP contains a lot of gray elements whose densities are neither zero nor one, we can only display the profile of the structure by using different values of the density threshold $\rho_{th}$. Actually, in our treatment, only the elements whose density values are greater than $\rho_{\text{th}}$ are plotted (i.e., $\rho > \rho_{\text{th}}$). Fig. 20a-Fig. 20c show the profiles of the optimized structure for $\rho_{th} = 0$, $\rho_{th} = 0.5$ and $\rho_{th} = 0.85$, respectively. It can be observed from these figures that the plotted structural profiles are highly dependent on the value of $\rho_{th}$ for this low value admissible volume fraction (i.e., 2%). Besides, it is not an easy task to transfer the optimization result to CAD/CAE systems for further treatment (note that the structure may be disconnected when a large $\rho_{th}$ is adopted while a small $\rho_{th}$ may lead to infeasible design). Some post-processing techniques are necessary to extract the structural profile from the gray image. However, it is also worth noting that the percentage of gray elements can be greatly reduced by enlarging the admissible volume of solid material or using some modern filter technique (Sigmund et al., 2016). For example, when the maximum admissible volume is chosen as $0.1V_D$, for different values of $\rho_{th}$, the corresponding optimized structures (obtained by the code of (Liu and Tovar, 2014)) are indeed very similar, as shown in Fig. 21. In addition, our numerical experiment also shows that by adopting the filter technique (Wang et al., 2011) in companion with a continuation process (filter radius is 2, threshold parameter is $\eta = 0.5$, and the projection steepness parameter $\beta$ is gradually increased from 0.5 to 64), as shown in Fig. 22, an almost black-and-white solution really can be obtained by the SIMP method with volume constraint $\bar{V} = 0.02V_D$, although some extra computational effort must be paid.

The entire structures obtained by the proposed method under three sets of background element meshes are shown in Fig. 23a-Fig. 23c, respectively. It can be observed that the optimized design obtained with a $42\times35\times42$ background mesh is almost an lattice-like structure, which is quite different from the ball-like structure shown in Fig.20a. This is due to the fact that since the minimum length scale in the optimized structures of MMC-based approach is limited by the characteristic size of the background mesh. Actually, for components with characteristic sizes less than the background mesh size, their contributions to structural stiffness cannot be detected by numerical integration procedure in FEA. Therefore, when the available material volume fraction is relatively small and the background mesh is not fine enough, it is extremely difficult to form a ball-like structure with very small thickness since the material distribution in MMC-based solution framework is purely black-and-white! Under this circumstance, only a lattice-like structure shown in Fig. 23a is selected to transmit the applied torque in a mechanically efficient way. Interestingly, by using the same FEA strategy in SIMP method to reanalyze the 1/8 structure of Fig. 23a, the compliance value is 126.62, which is very close to the result of SIMP approach. Of course, this problem can be well-addressed by adopting the adaptive mesh for FEA since we have explicit geometry description in the MMC-based approach. For the limitation of space, however, this issue will not be addressed in the present work. Furthermore, as shown in Fig. 23b and Fig. 23c, as the background element mesh is refined, the corresponding optimized structure gradually changes to a ball-like structure with more material distributing around the area where the external forces are applied. In addition, for the case where $\bar{V}=0.1V_\mathrm{D}$, by using a $126\times105\times126$ background element mesh for geometry description and $42\times35\times42$ hyper-elements for FEA, a closed sphere like structure can be obtained successfully (see Fig. 24). Furthermore, by taking the advantage of the explicit geometric description of the components, the optimized results can be directly transferred to CAD/CAE systems without any post-processing. The final optimized design displayed in CAD system is shown in Fig. 25.

In order to more accurately investigate the performances of the optimized designs with fine structural features obtained by the proposed approach, we transferred the 1/8 structures of Fig. 23a-Fig. 23c to Abaqus directly (thanks again to the explicit nature of geometry description in the MMC-based approach) and perform the FEA with a set of $126\times105\times126$ meshes. It is found that the corresponding values of structural compliance are 297.97, 196.40 and 197.58 respectively, which reveals that better designs do can be obtained by increasing the resolution of background element mesh. It is also worth noting that a direct comparison of the computational time between the proposed approach and the SIMP approach is not made for this example, since different optimizers are adopted for numerical optimization (i.e., OC method for the SIMP approach and MMA method for the proposed approach). However, since the number of design variables are only $720\times9=6480$ in the proposed approach while about 62000 in the SIMP-based approach, it can be expected that the computational time for updating design variables with the MMC approach will be much less than that of the SIMP approach if the same MMA optimizer is adopted. Some representative iteration curves are plotted in Fig. 26, respectively.

## 6. Concluding remarks

In the present work, a highly efficient MMC-based approach for multi-resolution topology optimization is proposed. With the use of this approach, both the numbers of the DOFs for finite element analysis and design variables for design optimization can be reduced substantially. Comparing with the traditional approaches, the corresponding computational time for the solution of large-scale topology optimization problems can be saved by about one order of magnitude. Compared to other based multi-resolution topology optimization methods, the proposed MMC-based multi-resolution method can generate optimized results with clearer boundaries and higher-resolution structural features with the use of linear finite elements more efficiently, and the optimized designs can be directly transferred to CAD/CAE systems without any post-processing. All these advantages can be attributed to the explicit nature of geometry description in the MMC-based solution framework. As preliminary attempt, only minimum compliance design problems are considered in the present study to demonstrate the effectiveness of the proposed approach. It can be expected that the proposed approach can also find applications in other computationally intensive optimization problems (e.g., structural optimization considering geometry/material nonlinearity). Another interesting research direction is combining both the advantages of the implicit SIMP-based approaches and the explicit MMC-based approaches to develop some hybrid approaches for solving topology optimization problems where more complicate objective/constraint functions/functionals are involved. This is highly possible since as discussed at the end of Section 4, the proposed solution framework is general enough to achieve this goal. Corresponding research results will be reported elsewhere.

# Appendix

The process of generating TDF locally can be elaborated as follows:

1) Generating a rectangle $\Omega_i^{\text{ext}}$ (pink region), with the use of the parameters $(\boldsymbol{o}_i, \theta_i, l_i, t_i)$, as shown in Fig. A1b. Here the symbol $\boldsymbol{o}_i = (x_{i0}, y_{i0})^\top$ is the vector of the coordinates of the central point of the component, $\theta_i$ is the corresponding inclined angle, while $l_i = 2a_i \sqrt[6]{(1+\epsilon)}$ and $t_i = \max(2t_i^1, 2t_i^2) \sqrt[6]{(1+\epsilon)}$ are the length and width of $\Omega_i^{\text{ext}}$, respectively. Note that $\Omega_i^{\text{ext}} \supset \Omega_i' = \{\boldsymbol{x} | \boldsymbol{x} \in D, \phi_i(\boldsymbol{x}) \geq -\epsilon\}$ (yellow part).

2) From the vertexes (which can be found analytically) of $\Omega_i^{\text{ext}}$, generating another rectangle $\Omega_i^{\text{rec}}$ (light blue region), as shown in Fig. A1b.

3) Generating the TDF associated with $\Omega_i^{\text{rec}}$ (this can be done very easily).

4) Finding the TDF values in $\Omega_i^{\text{rec}}$ such that $-\epsilon \leq \phi_i(\boldsymbol{x}) \leq \epsilon$ and only storing these values by *sparse matrix* for subsequent treatment.

The above treatment guarantees that only local values of $\phi_i(\boldsymbol{x})$ are evaluated in the corresponding manipulations, which saves the computational effort substantially.

25
**Acknowledgements**

The authors would like to thank Prof. Oded Amir for constructive discussions and the valuable comments from anonymous reviewers on improving the quality of the present work. The financial supports from the National Key Research and Development Plan (2016YFB0201600, 2016YFB0201601, 2017YFB0202800, 2017YFB0202802), the National Natural Science Foundation (11402048, 11472065, 11732004, 11772026, 11772076), Program for Changjiang Scholars, Innovative Research Team in University (PCSIRT) and 111 Project (B14013) are also gratefully acknowledged.

# Tables

Table. 1 Optimization results of the cantilever beam example obtained by the proposed approach with different FE meshes.

| Optimized structure / Performances | Number of FE mesh | $\bar{t}_{\text{TDF}}$ (s) | $\bar{t}_{\text{FEA}}$ (s) | $\bar{t}_{\text{sen}}$ (s) | $\bar{t}_{\text{MMA}}$ (s) | $\bar{t}_{\text{total}}$ (s) | $n_{\text{iter}}$ | $c_{\text{obj}}$ | $c_{\text{post}}$ | Relative FEA error $\left(\dfrac{|c_{\text{post}} - c_{\text{obj}}|}{c_{\text{post}}}\right)$ |
|---|---|---|---|---|---|---|---|---|---|---|
| 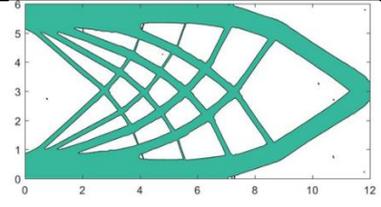 | 1280×640 | 1.47 | 15.18 | 0.15 | 0.06 | 18.62 | 96 | 73.60 | 73.60 | 0.00% |
| 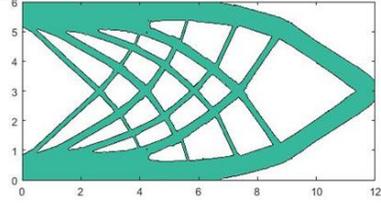 | 640×320 | 1.19 | 6.54 | 0.12 | 0.05 | 9.69 | 121 | 73.61 | 73.99 | 0.51% |
| 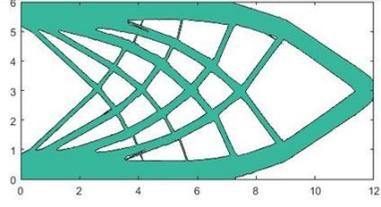 | 320×160 | 1.28 | 1.61 | 0.13 | 0.05 | 4.78 | 161 | 72.76 | 73.84 | 1.46% |

| | | | | | | | | | |
|---|---|---|---|---|---|---|---|---|---|
| 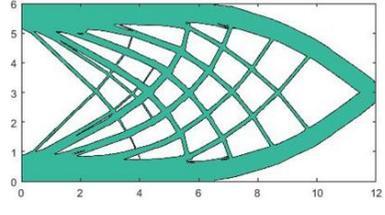 | 256×128 | 1.25 | 1.25 | 0.13 | 0.05 | 4.29 | 154 | 72.16 | 73.70 | 2.09% |
| 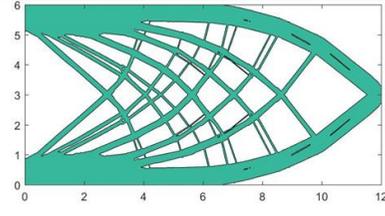 | 160×80 | 0.81 | 0.47 | 0.10 | 0.03 | 3.13 | 124 | 71.30 | 73.73 | 3.30% |
| 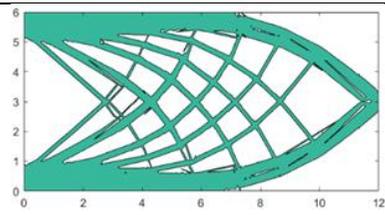 | 128×64 | 0.89 | 0.36 | 0.11 | 0.03 | 3.02 | 188 | 71.18 | 79.85 | 10.86% |
| 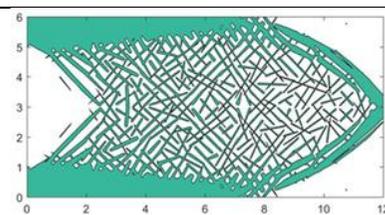 | 64×32 | 0.89 | 0.25 | 0.10 | 0.04 | 2.93 | / | 56.53 | 1010.09 | 94.40% |

Table. 2 Optimization results of the cantilever beam example obtained with the SIMP approach.

| Performances<br>Optimized structure | Number of FE mesh | $\bar{t}_{\text{FEA}}$ (s) | $\bar{t}_{\text{sen}}$ (s) | $\bar{t}_{\text{MMA}}$ (s) | $\bar{t}_{\text{total}}$ (s) | $n_{\text{iter}}$ | $c_{\text{obj}}$ |
|---|---|---|---|---|---|---|---|
| 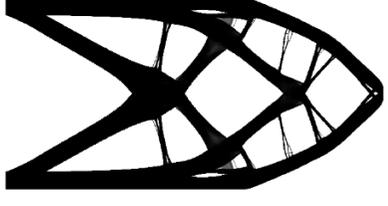 | 1280×640 | 13.67 | 0.19 | 14.70 | 28.91 | 251 | 80.29 |

Table. 3 Optimization results of the MBB example obtained with the proposed approach under different FE meshes.

| Optimized structure / Performance | Number of FE mesh | $\bar{t}_{\text{TDF}}$ (s) | $\bar{t}_{\text{FEA}}$ (s) | $\bar{t}_{\text{sen}}$ (s) | $\bar{t}_{\text{MMA}}$ (s) | $\bar{t}_{\text{total}}$ (s) | $n_{\text{iter}}$ | $c_{\text{obj}}$ | $c_{\text{post}}$ | Relative FEA error $\left(\dfrac{|c_{\text{post}} - c_{\text{obj}}|}{c_{\text{post}}}\right)$ |
|---|---|---|---|---|---|---|---|---|---|---|
| 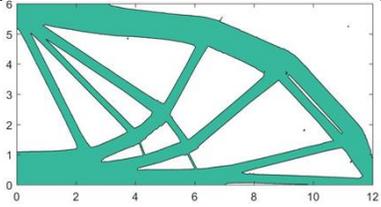 | 1280×640 | 1.67 | 14.79 | 0.18 | 0.05 | 18.34 | 117 | 96.59 | 96.59 | 0.00% |
| 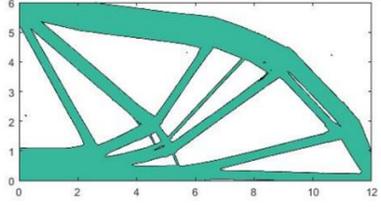 | 640×320 | 1.77 | 6.56 | 0.19 | 0.06 | 10.22 | 144 | 94.78 | 96.70 | 1.99% |
| 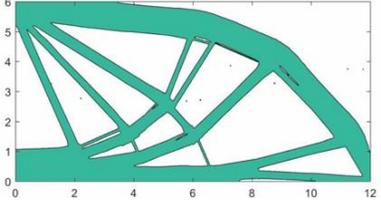 | 320×160 | 1.88 | 1.60 | 0.19 | 0.07 | 5.38 | 191 | 91.43 | 96.25 | 5.01% |

| 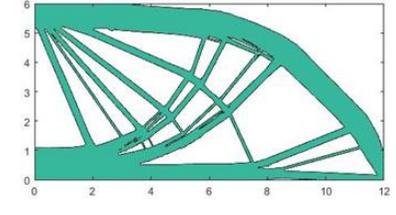 | 256×128 | 1.56 | 1.06 | 0.16 | 0.05 | 4.52 | 224 | 90.32 | 96.98 | 6.87% |
| 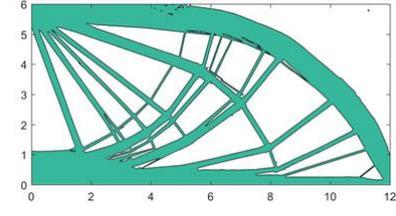 | 160×80 | 1.35 | 0.59 | 0.14 | 0.05 | 3.86 | 202 | 87.87 | 97.86 | 10.21% |

Table. 4 Optimization results of the cantilever beam example obtained with the SIMP approach.

| Performance / Optimized structure | Number of FE mesh | $\bar{t}_{\text{FEA}}$ (s) | $\bar{t}_{\text{sen}}$ (s) | $\bar{t}_{\text{MMA}}$ (s) | $\bar{t}_{\text{total}}$ (s) | $n_{\text{iter}}$ | $c_{\text{obj}}$ |
|---|---|---|---|---|---|---|---|
| 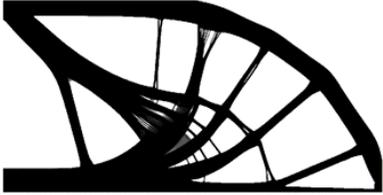 | 1280×640 | 11.78 | 0.20 | 12.51 | 24.80 | 245 | 106.23 |

Table. 5 Optimization results of the example under distributed load obtained by the proposed approach and different FE meshes.

| Performances / Optimized structure | Number of FE mesh | $\bar{t}_{\text{TDF}}$ (s) | $\bar{t}_{\text{FEA}}$ (s) | $\bar{t}_{\text{sen}}$ (s) | $\bar{t}_{\text{MMA}}$ (s) | $\bar{t}_{\text{total}}$ (s) | $n_{\text{iter}}$ | $c_{\text{obj}}$ | $c_{\text{post}}$ | Relative FEA error $\left(\dfrac{|c_{\text{post}} - c_{\text{obj}}|}{c_{\text{post}}}\right)$ |
|---|---|---|---|---|---|---|---|---|---|---|
| 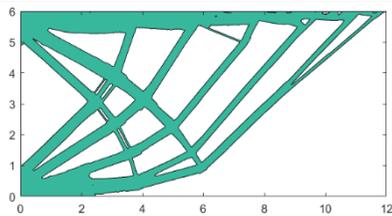 | 1200×600 | 0.86 | 12.71 | 0.08 | 0.04 | 15.15 | 131 | 17.63 | 17.63 | - |
| 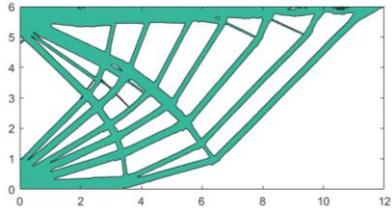 | 600×300 | 1.05 | 5.81 | 0.11 | 0.03 | 8.56 | 200 | 17.36 | 17.55 | 1.10% |
| 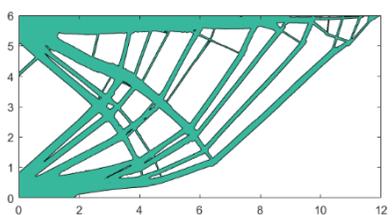 | 400×200 | 1.37 | 2.49 | 0.12 | 0.06 | 5.63 | 187 | 17.14 | 18.27 | 6.20% |

| | | | | | | | | | |
|---|---|---|---|---|---|---|---|---|---|
| 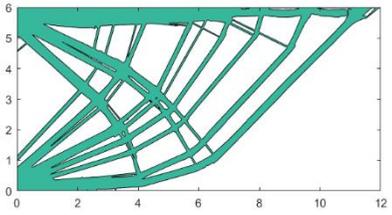 | 300×150 | 1.26 | 1.39 | 0.12 | 0.06 | 4.39 | 159 | 17.11 | 32.49 | 47.34% |
| 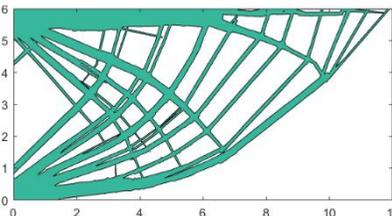 | 200×100 | 1.33 | 0.67 | 0.13 | 0.07 | 3.85 | 182 | 16.69 | 26.09 | 36.03% |
| 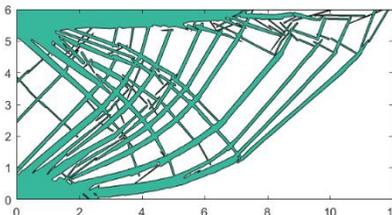 | 120×60 | 1.27 | 0.46 | 0.16 | 0.05 | 3.72 | / | 16.17 | 939.21 | 98.28% |

# Figures

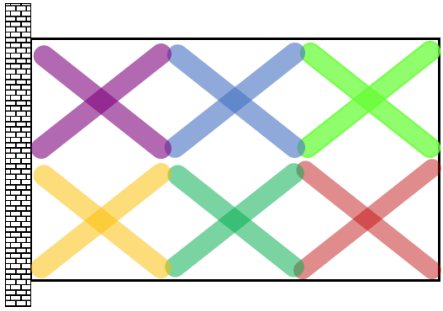

(a) The initial layout of the components.

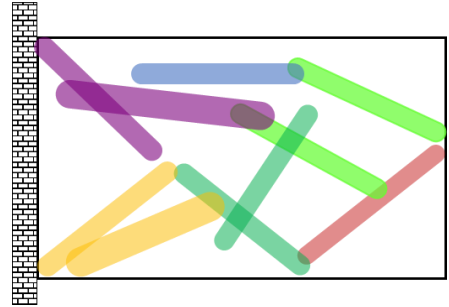

(b) Optimization process.

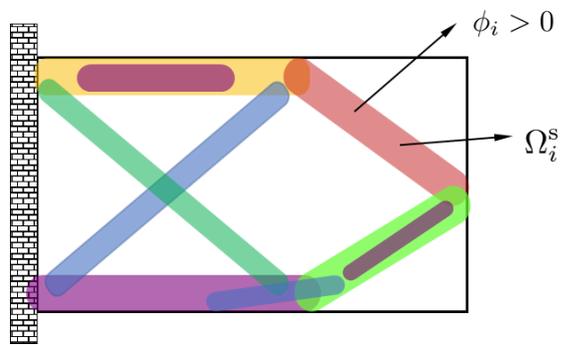

(c) The optimized layout of the components.

Fig. 1 A schematic illustration of the MMC-based topology optimization method.

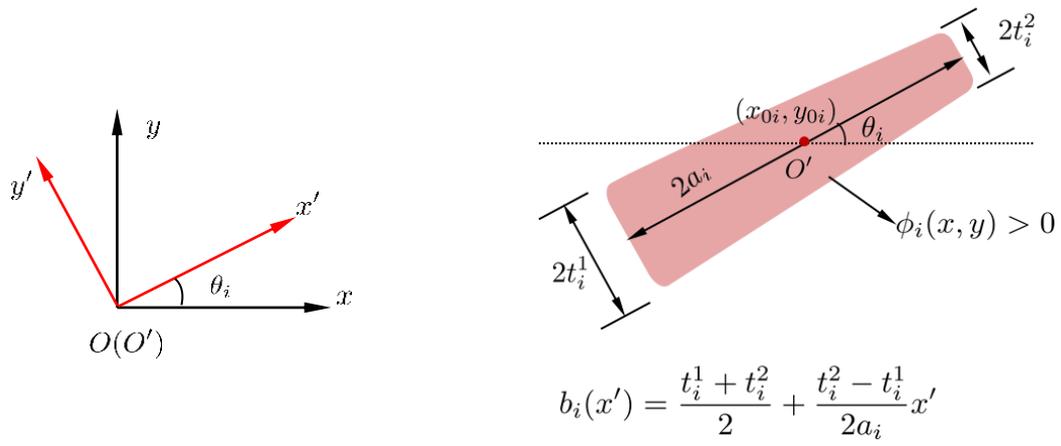

Fig. 2 The geometry description of a two-dimensional structural component.

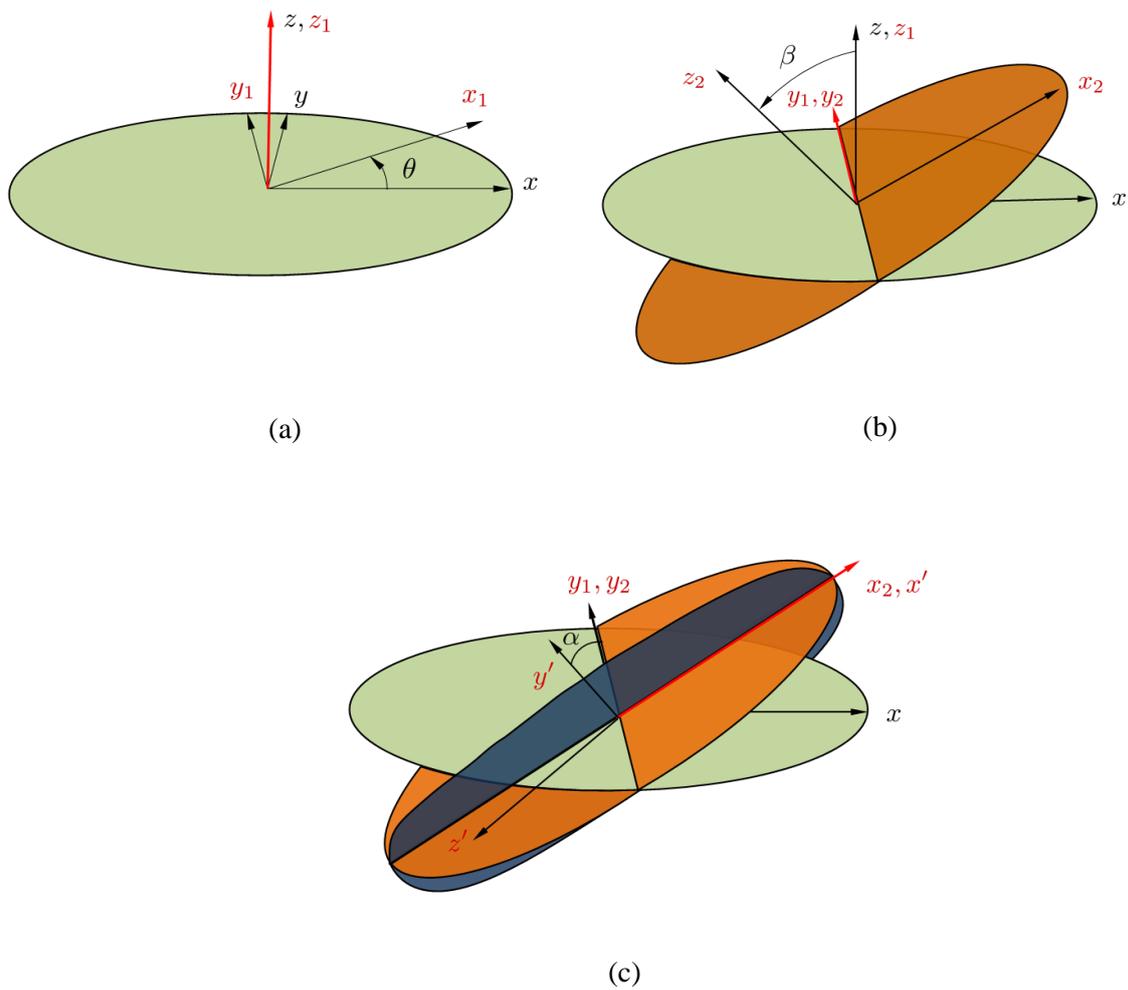

Fig. 3 Coordinate transformation associated with a three-dimensional component.

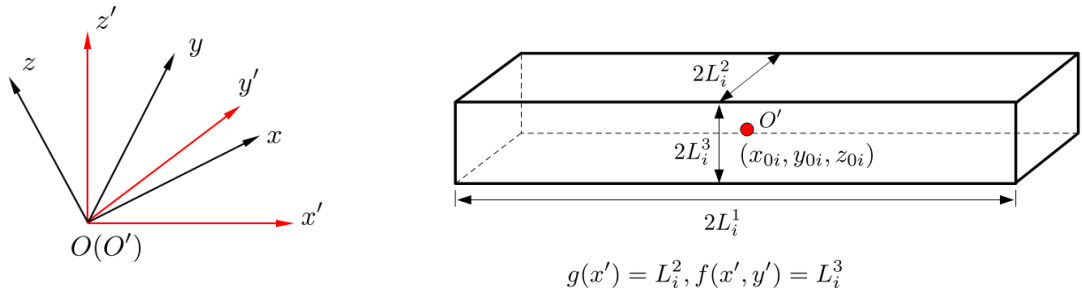

Fig. 4 The geometry description of a three-dimensional structural component.

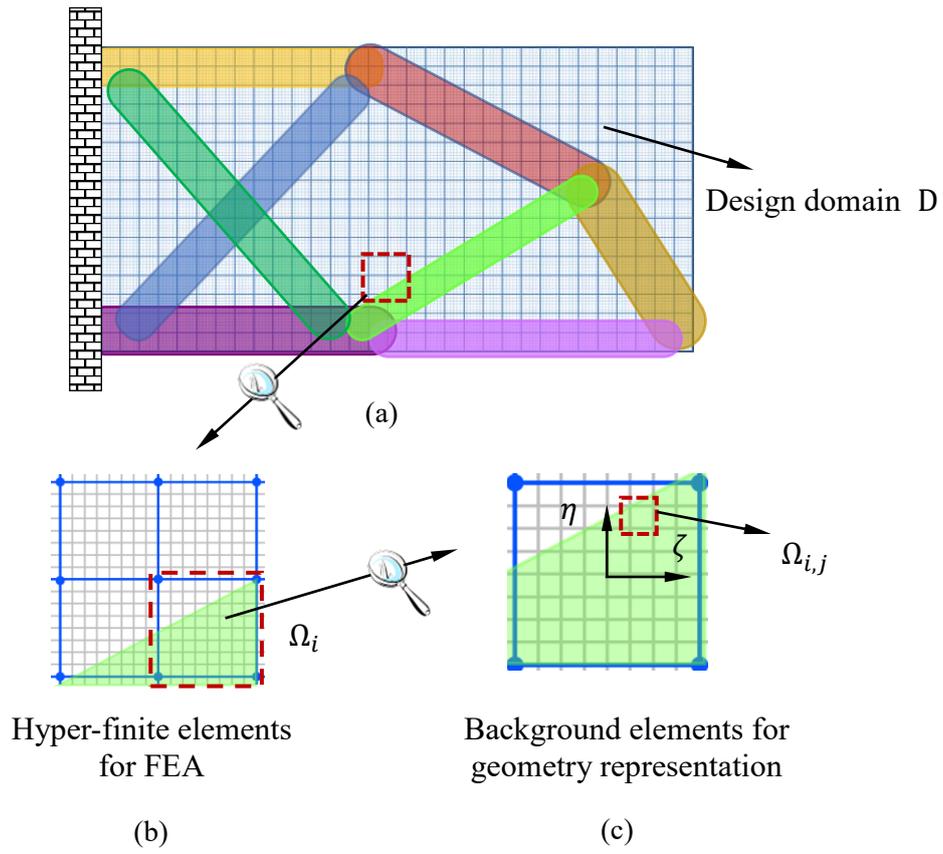

Fig. 5 A schematic illustration of the basic idea of the proposed MMC-based multi-resolution topology optimization approach.

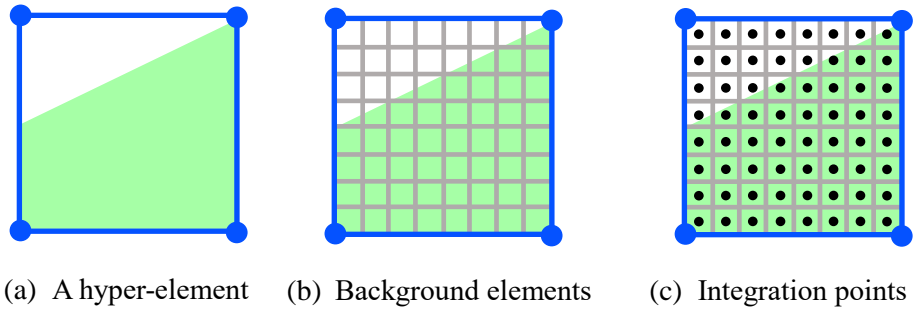

(a) A hyper-element    (b) Background elements    (c) Integration points

Fig. 6 A schematic illustration of a hyper-element, the corresponding background elements and integration points.

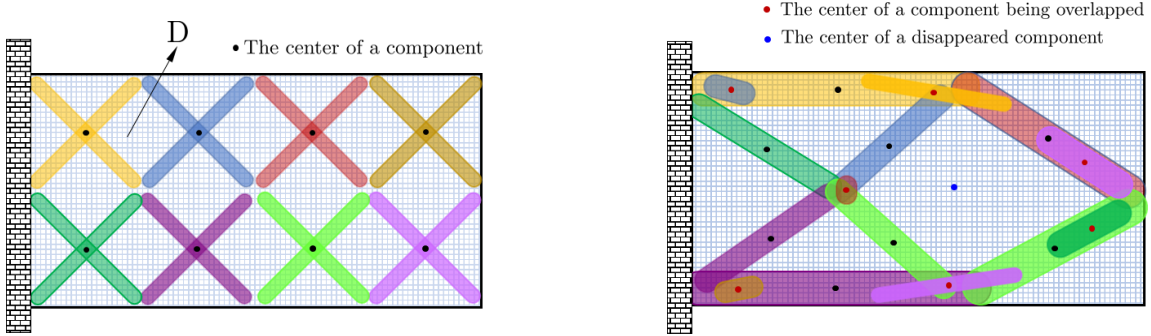

(a) MMC-based topology optimization without design domain partitioning.

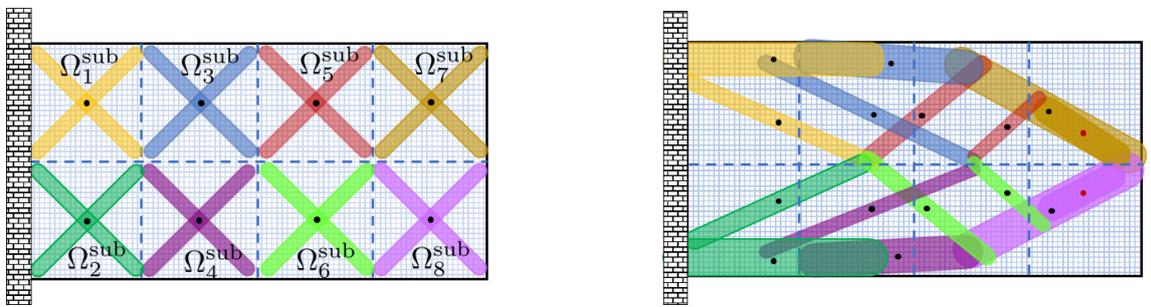

(b) MMC-based topology optimization with design domain partitioning.

Fig. 7 The basic idea of the design domain partitioning strategy.

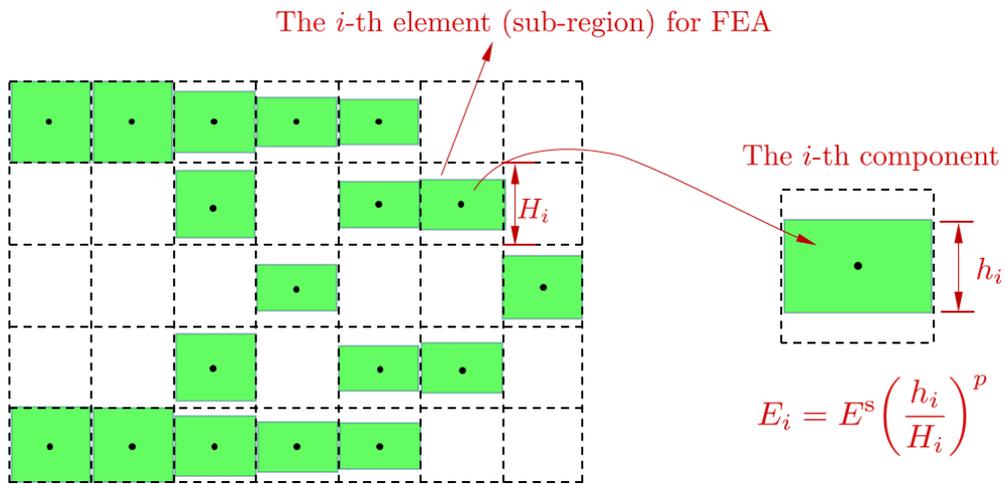

Fig. 8 The degeneration of the MMC-based approach to the SIMP approach.

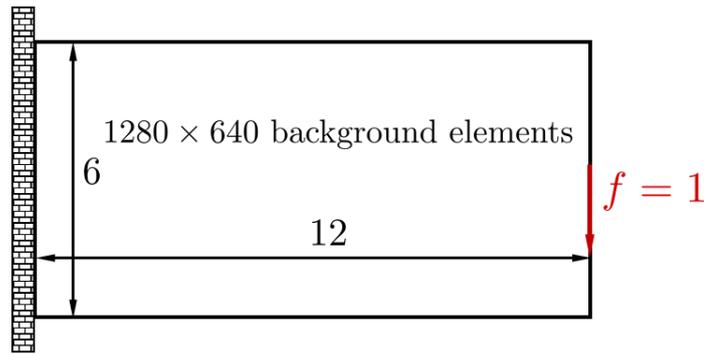

Fig. 9 The cantilever beam example.

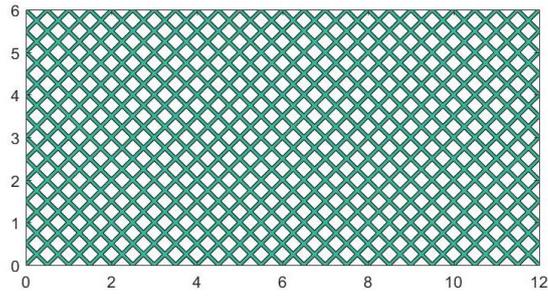

Fig. 10 The initial design of the cantilever beam example.

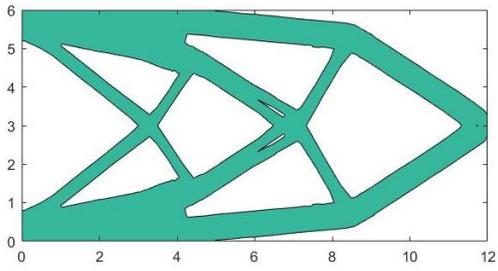
(a) Compliance value 74.63.

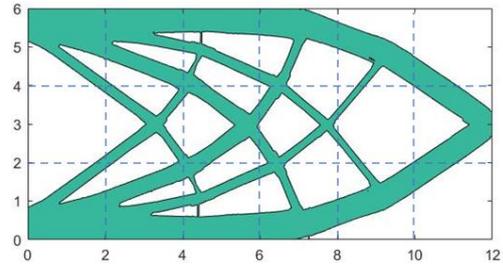
(b) Compliance value 73.60.

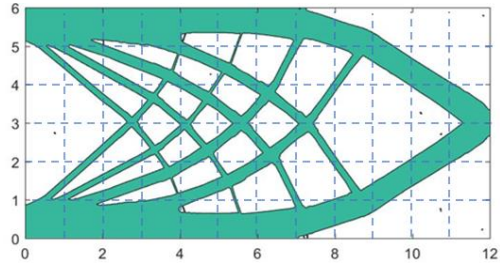
(c) Compliance value 73.59.

Fig. 11 The optimized structures obtained with the design domain partitioning strategy.

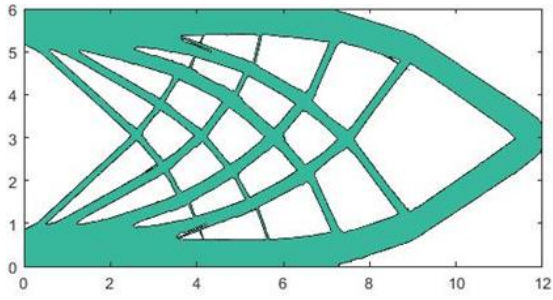 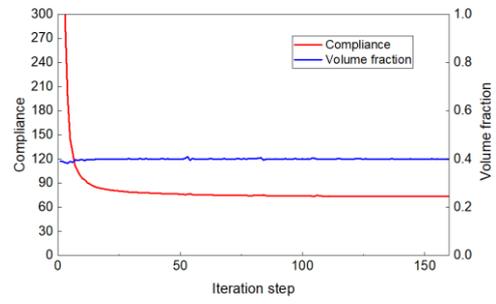

(a) 320×160 hyper elements.

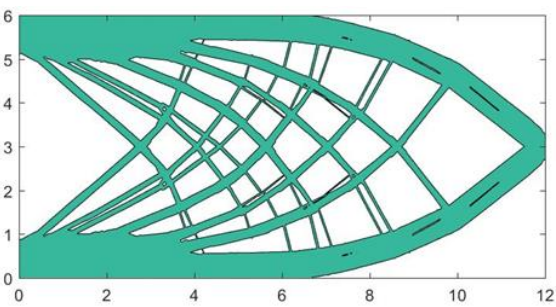 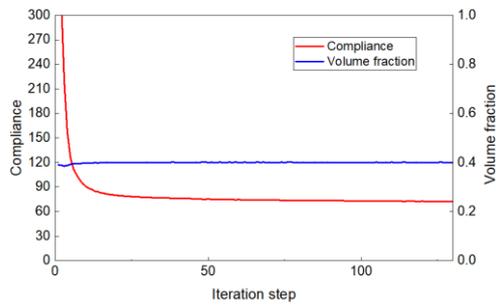

(b) 160×80 hyper elements.

Fig. 12 The optimized structures and the corresponding iteration curves of the cantilever beam example obtained by the proposed approach.

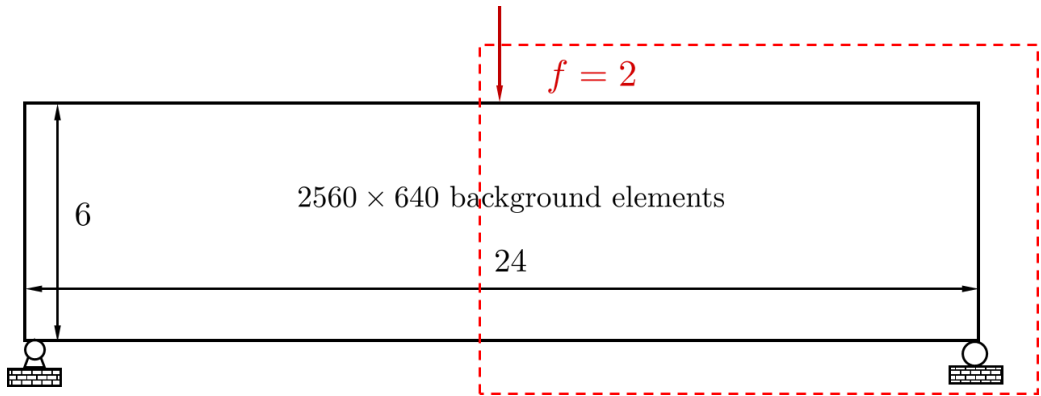

(a)

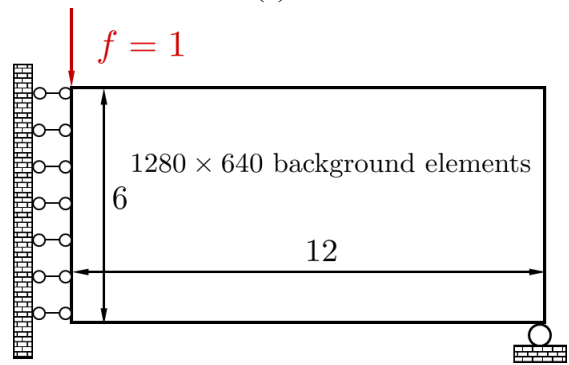

(b)

Fig. 13 The MBB example.

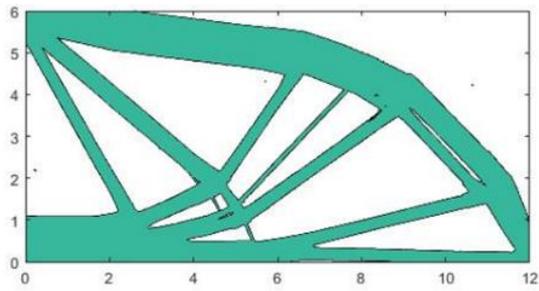 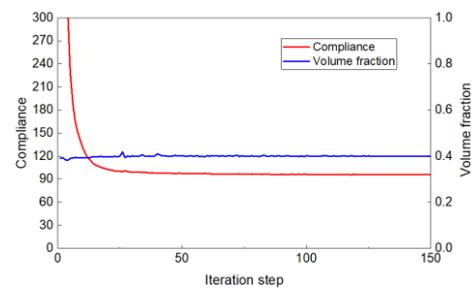

(a) 640×320 hyper elements.

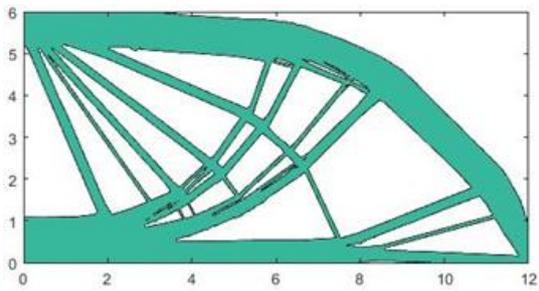 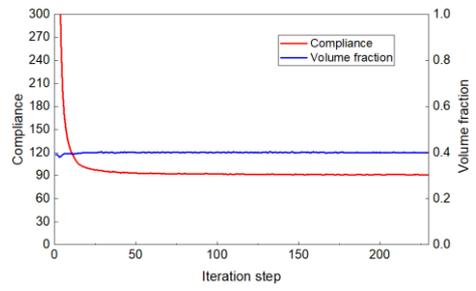

(b) 256×128 hyper elements.

Fig. 14 The optimized structures and the corresponding iteration curves of the MBB example obtained by the proposed approach.

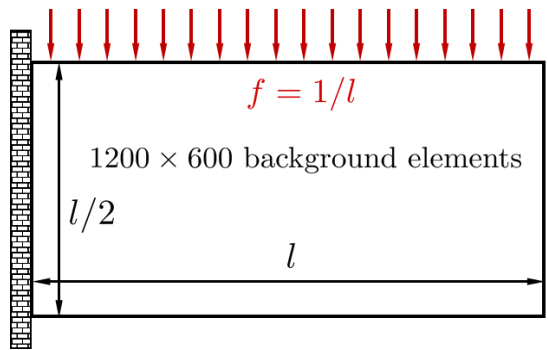

Fig. 15 A cantilever beam example under uniformly distributed load.

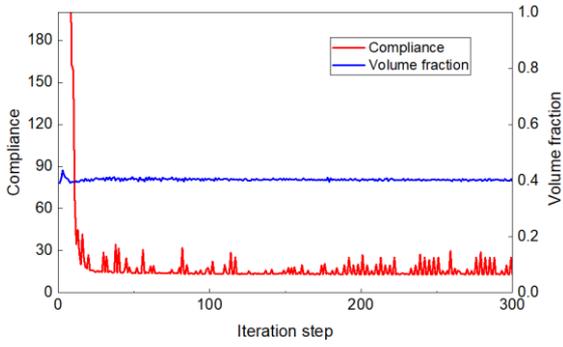
(a) The iteration history.

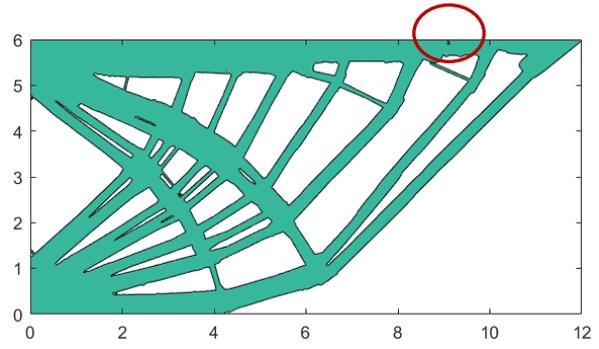
(b) 151 step, compliance 14.37.

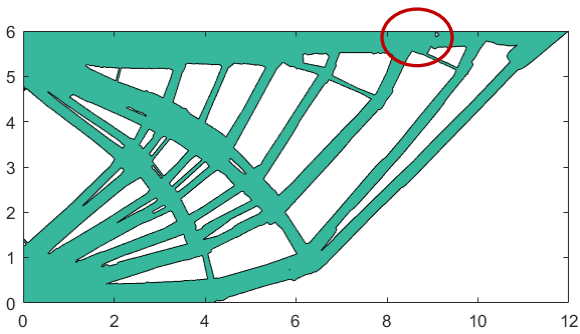
(c) 152 step, compliance 17.63.

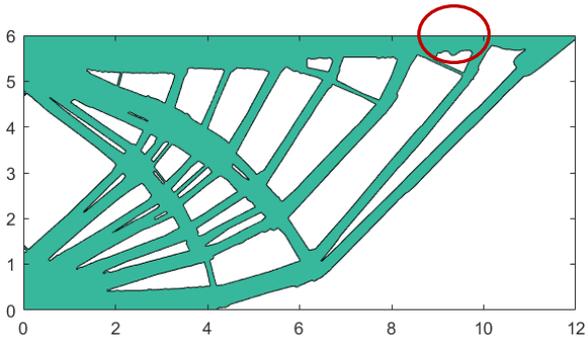
(d) 153 step, compliance 13.25.

Fig. 16 The results of the cantilever beam example under uniformly distributed load obtained by the proposed approach ($\bar{V} = 0.4V_\mathrm{D}$ and 1200×600 FE mesh).

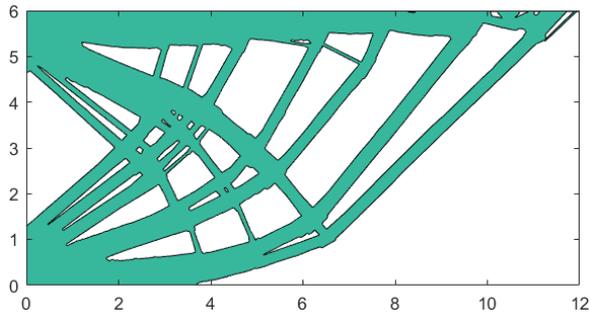 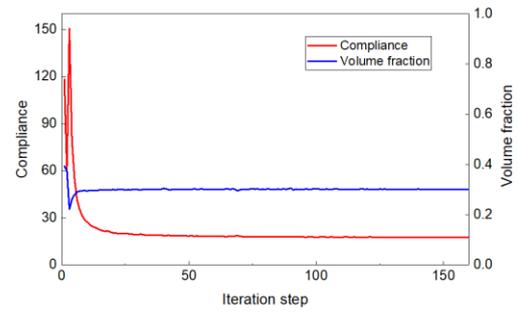

(a) The optimized structure with $c_{\text{obj}} = 13.16$.

(b) The iteration history.

Fig. 17 The results of the cantilever beam example under uniformly distributed load obtained by the proposed approach ($\bar{V} = 0.4V_{\text{D}}$, 1200 ×600 FE mesh, and one layer of fixed solid region).

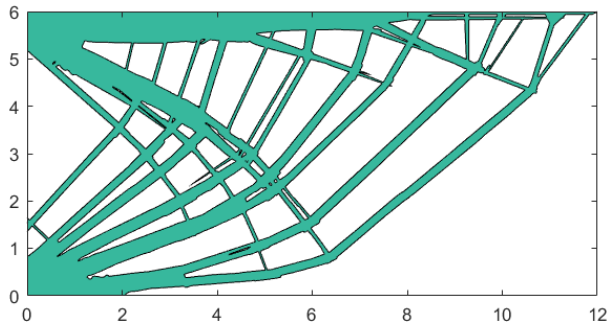 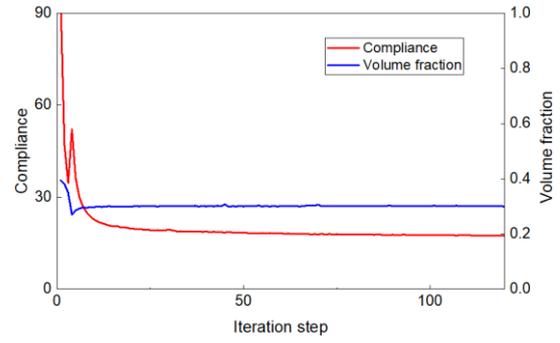

(a) 300×150 hyper-elements ($n_{\text{iter}} = 114$ and $c_{\text{obj}} = 16.99$, $c_{\text{post}} = 18.09$, relative FEA error 6.08%).

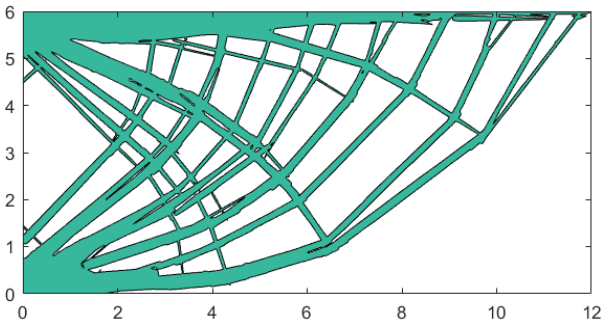 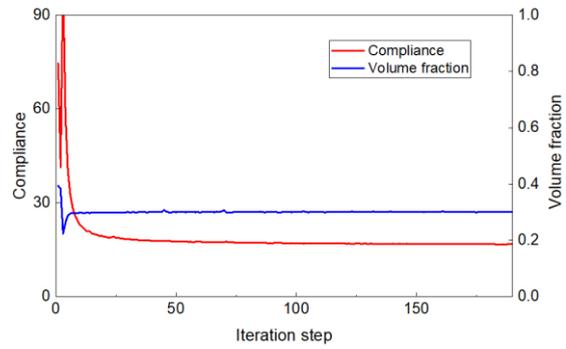

(b) 200×100 hyper-elements ($n_{\text{iter}} = 185$ and $c_{\text{obj}} = 16.68$, $c_{\text{post}} = 17.83$, relative FEA error 6.45%).

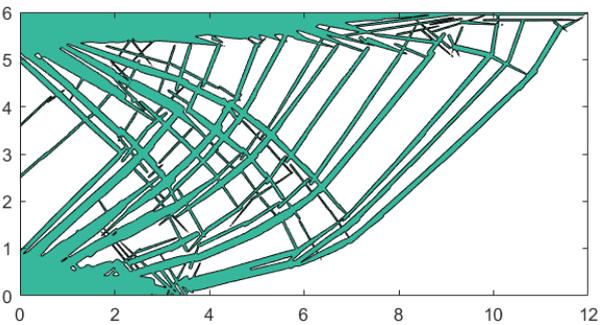 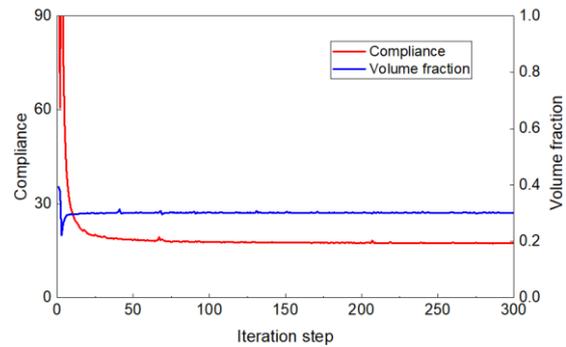

(c) 120×60 hyper-elements ($c_{\text{obj}} = 16.31$, $c_{\text{post}} = 357.46$, relative FEA error 95.44%).

Fig. 18 The results of the cantilever beam example under uniformly distributed load obtained by the proposed approach ($\bar{V} = 0.3V_{\text{D}}$, 1200 ×600 background elements and six layers of fixed solid region).

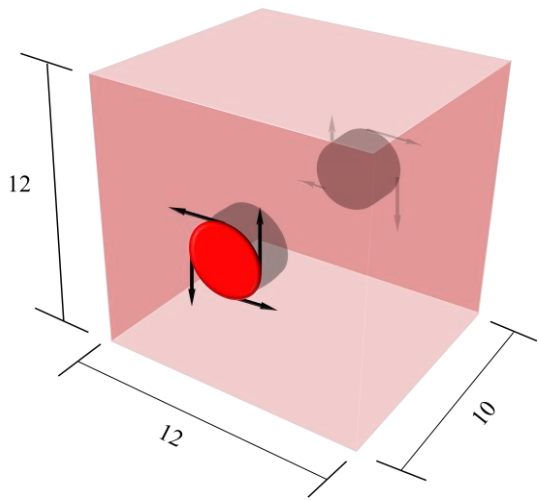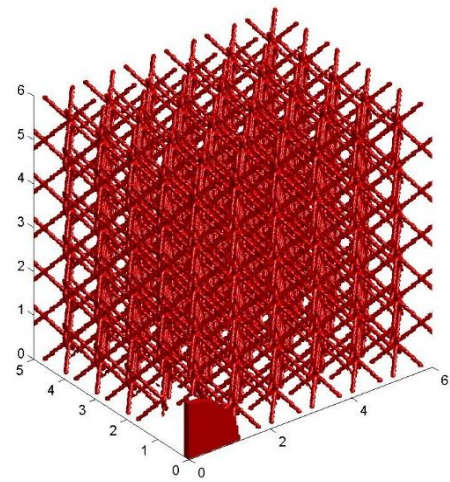

(a) The design domain.   (b) The initial design.

Fig. 19 The 3D box example.

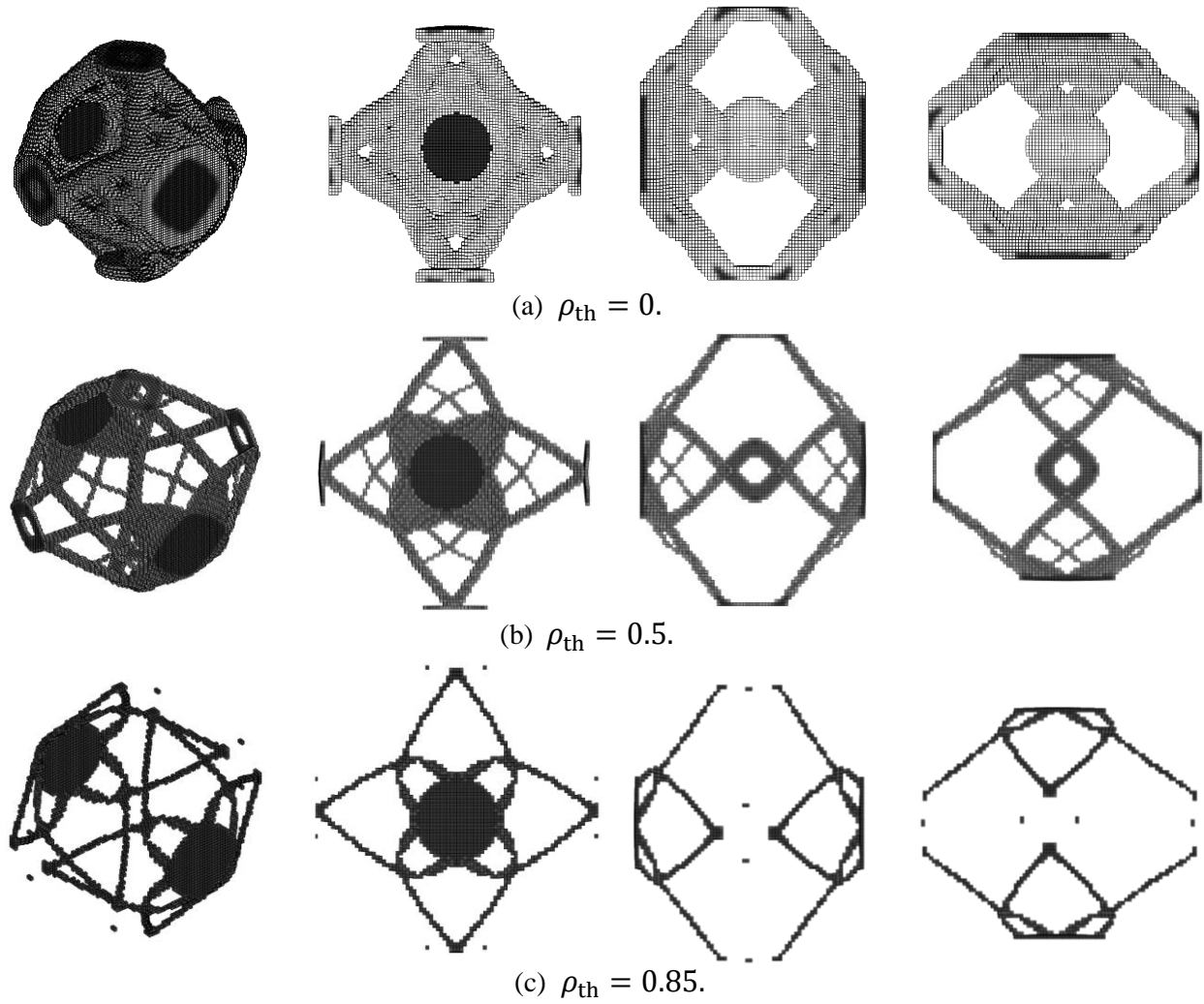

(a) $\rho_{th} = 0$.

(b) $\rho_{th} = 0.5$.

(c) $\rho_{th} = 0.85$.

Fig. 20 The optimized structure of the 3D box example obtained with the SIMP method (displayed with different values of $\rho_{th}$) at $n_{iter}$=108.

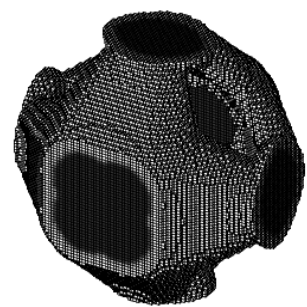 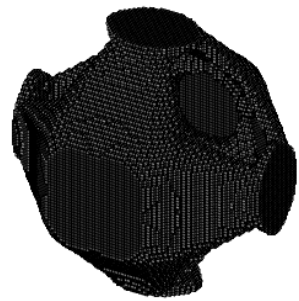 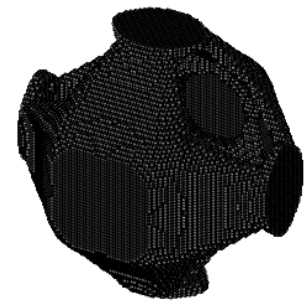

(b) $\rho_{th} = 0.$    (c) $\rho_{th} = 0.5.$    (a) $\rho_{th} = 0.85.$

Fig. 21 The optimized structures of the 3D box example by SIMP method ($\bar{V} = 0.1V_D$ and 42×35 ×42 FE mesh) at $n_{iter}$=35.

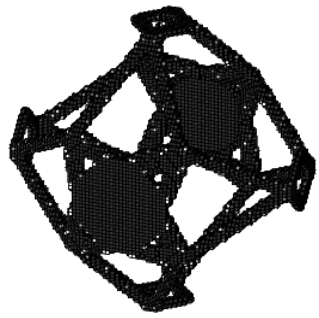 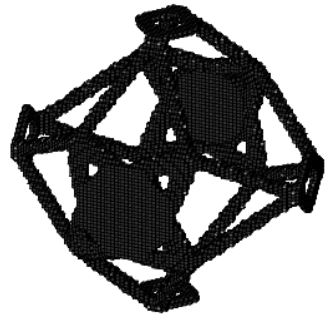 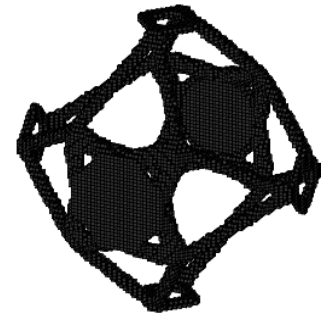

(a) $\rho_{th} = 0.001$.     (b) $\rho_{th} = 0.5$.     (c) $\rho_{th} = 0.85$.

Fig. 22 The optimized structures of the 3D box example by SIMP method via adopting both density filter and threshold projection techniques ($\bar{V} = 0.02V_D$ and 42×35 ×42 FE mesh) at $n_{iter}$=128.

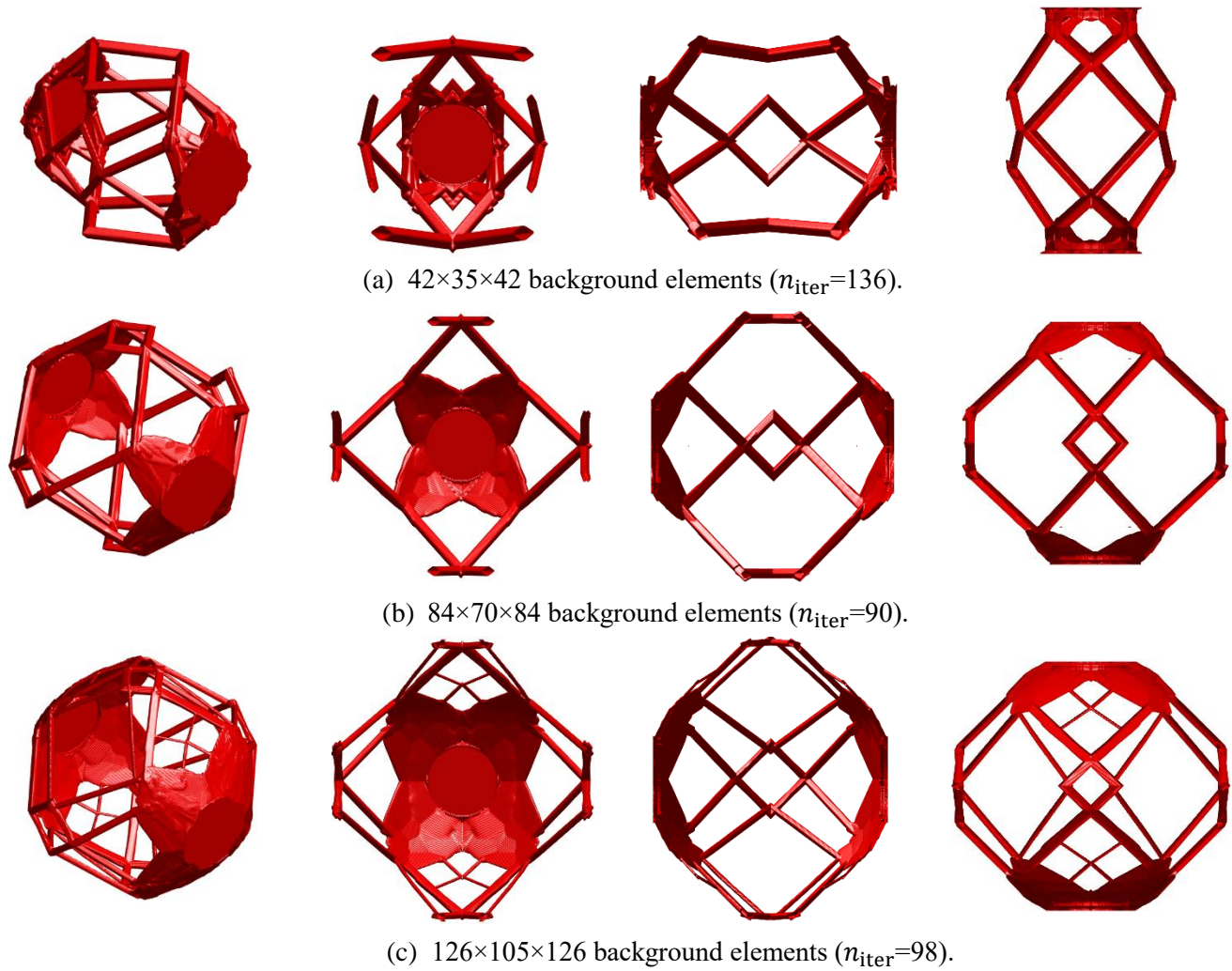

(a) 42×35×42 background elements ($n_{iter}$=136).

(b) 84×70×84 background elements ($n_{iter}$=90).

(c) 126×105×126 background elements ($n_{iter}$=98).

Fig. 23 The optimized structures of the 3D box example obtained with 42×35 ×42 hyper-elements and different background elements by the proposed method.

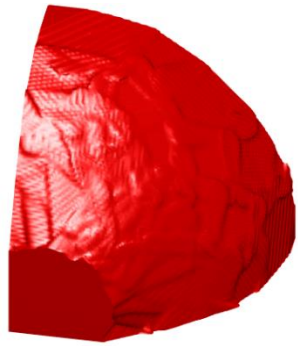 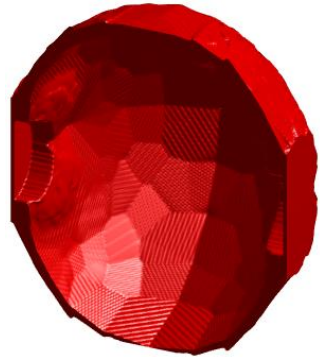 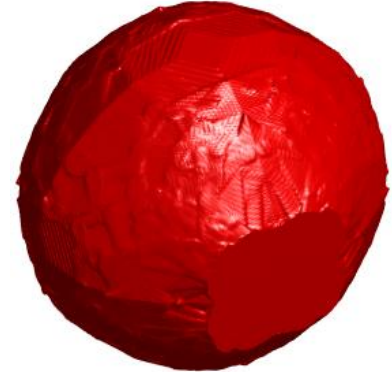

(a) 1/8 optimized structure.     (b) 1/2 optimized structure.     (c) The full optimized structure.

Fig. 24 The optimized structures of the 3D box example by the proposed method ($\bar{V} = 0.1V_\mathrm{D}$, 126×105×126 background elements and 42×35×42 hyper-elements) at $n_\mathrm{iter}$=116.

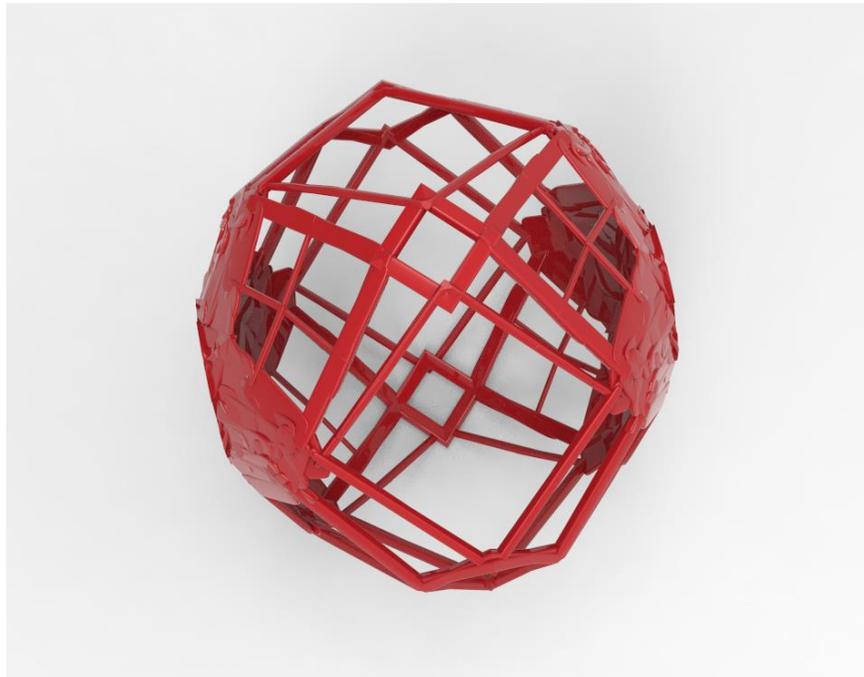

Fig. 25 The CAD model of the optimized structure obtained by the proposed method ($\bar{V} = 0.02V_\mathrm{D}$, 126×105×126 background elements and 42×35×42 hyper-elements).

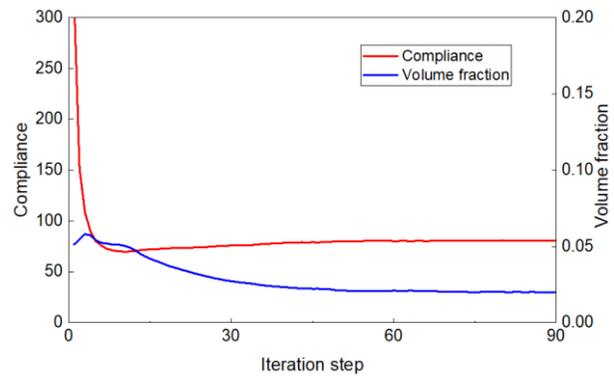 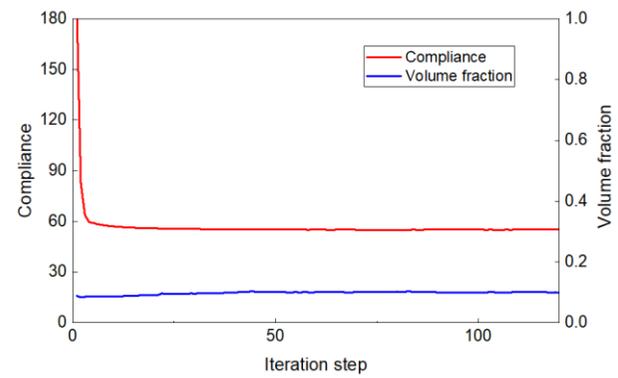

(a) $\bar{V} = 0.02V_D$, 84×70×84 background elements and 42×35×42 hyper-elements.

(b) $\bar{V} = 0.1V_D$, 126×105×126 background elements and 42×35×42 hyper-elements.

Fig. 26 The iteration curves of two considered cases using the proposed method.

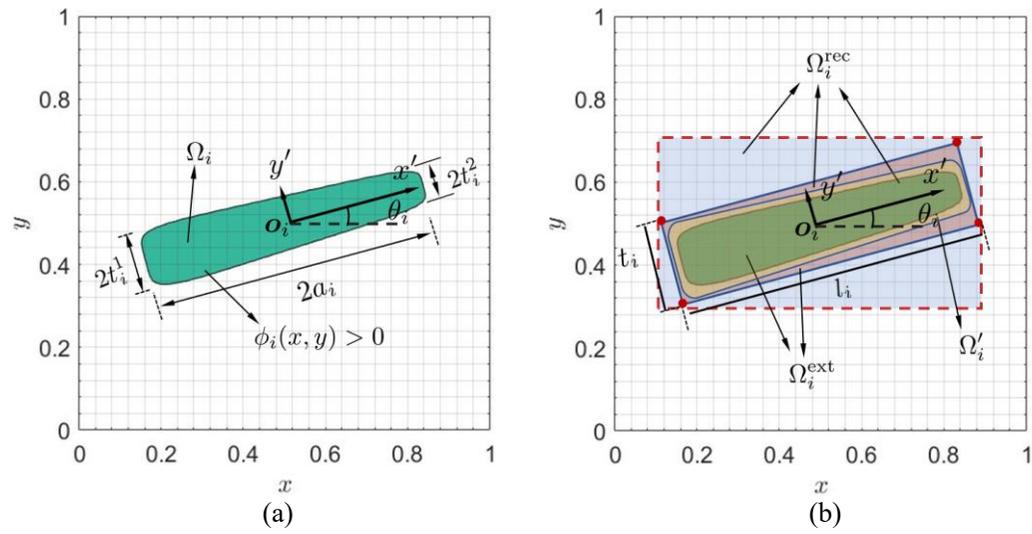

Fig. A1 A schematical illustration of generating the TDF locally..